\documentclass[12pt,draftclsnofoot,onecolumn]{IEEEtran}

\usepackage[T1]{fontenc}
\usepackage{xspace,amsmath,amssymb,amsfonts,epsfig,syntonly,bbm,dsfont}
\usepackage{cite,bm,color,url,textcomp,empheq,boxedminipage}
\usepackage{algorithmicx,algorithm,algpseudocode}
\usepackage{epstopdf}
\usepackage{empheq}
\usepackage{graphicx,graphics,subfigure}
\usepackage{multirow,multicol}
\usepackage{psfrag}
\usepackage{stfloats}

\newtheorem{lemma}{\underline{Lemma}}[section]

\newtheorem{proposition}{\underline{Proposition}}[section]

\DeclareMathOperator*{\argmax}{arg\,max}

\long\def\symbolfootnote[#1]#2{\begingroup
\def\thefootnote{\fnsymbol{footnote}}
\footnote[#1]{#2}\endgroup}

\psfull

\begin{document}

\title{Amplify-and-Forward Relaying for Hierarchical Over-the-Air Computation}
\author{Feng Wang, Jie Xu, Vincent K. N. Lau, and Shuguang Cui\\
\thanks{This paper was presented in part at the IEEE Global Communications Conference Workshop on Edge Learning over 5G Networks and Beyond, Taipei, China, Dec. 2020\cite{conf-version}.}

\thanks{F. Wang is with the School of Information Engineering, Guangdong University of Technology, Guangzhou 510006, China, and also with the Department of Electronic and Computer Engineering, The Hong Kong University of Science and Technology, Hong Kong (e-mail: fengwang13@gdut.edu.cn).}

\thanks{J. Xu and S. Cui are with the School of Science and Engineering and Future Network of Intelligence Institute (FNii), The Chinese University of Hong Kong (Shenzhen), Shenzhen 518172, China, and also with Peng Cheng Laboratory, Shenzhen 518066, China (e-mail: xujie@cuhk.edu.cn, shuguangcui@cuhk.edu.cn). J. Xu is the corresponding author.}

\thanks{V. K. N. Lau is with the Department of Electronic and Computer Engineering, The Hong Kong University of Science and Technology, Hong Kong (e-mail: eeknlau@ust.hk).}

\vspace{-1.8cm}
}

\maketitle

\begin{abstract}
  Over-the-air computation (AirComp) has emerged as a promising technique in future intelligent wireless networks, which enables swift functional computation among distributed wireless devices (WDs) by exploiting the superposition property of wireless channels. This paper studies a new {\em hierarchical} AirComp network over a large area, in which a set of intermediate relays are exploited to facilitate the massive data aggregation from a large number of WDs. Under this setup, we present a two-phase amplify-and-forward (AF) relaying protocol. In the first phase, the WDs simultaneously send their data to the relays, while in the second phase, the relays amplify the respectively received signals and concurrently forward them to the fusion center (FC) for aggregation. Our objective is to minimize the computational mean squared error (MSE) at the FC, by jointly optimizing the transmit coefficients of the WDs, the AF coefficients of the relays, and the de-noising factor of the FC, subject to their individual transmit power constraints. First, we consider the centralized design with global channel state information (CSI), in which the inter-relay signals can be exploited beneficially for data aggregation. In this case, we develop an alternating-optimization-based algorithm to obtain a high-quality solution to the computational MSE minimization problem. The obtained solution shows that the phase of the transmit coefficient at each WD is opposite to that of the WD-relay-FC channel to ensure the signal phase alignment at the FC, and the transmit power of each WD/relay follows a regularized composite-channel-inversion structure to strike a balance between minimizing the signal-magnitude-misalignment-induced error and the noise-induced error. Next, to reduce the signaling overhead caused by the centralized design, we consider an alternative decentralized design with partial CSI, in which the relays and the FC make their own decisions by only requiring the channel power gain information across different relays. In this case, the relays and FC need to treat the inter-relay signals as harmful interference or noise. Accordingly, we optimize the transmit coefficients of the WDs associated with each relay, and the relay AF coefficients (together with the FC de-noising factor) in an iterative manner, which can be implemented efficiently in a decentralized way. Finally, numerical results show the fast convergence of the proposed centralized and decentralized designs. It is also shown that both designs achieve significant MSE performance gains over benchmark schemes without the joint optimization.
\end{abstract}

\vspace{-0.2cm}
\begin{IEEEkeywords}
\vspace{-0.2cm}
Hierarchical over-the-air computation (AirComp), amplify-and-forward relaying, mean squared error (MSE), optimization.
\end{IEEEkeywords}

\IEEEpeerreviewmaketitle


\section{Introduction}
 The recent advancements in beyond fifth-generation (B5G) and sixth-generation (6G) wireless networks and the proliferation of smart wireless devices (WDs) are expected to enable various intelligent applications such as auto-driving and virtual/agugmented realities. Towards this end, new techniques such as distributed sensing \cite{Fan21} and distributed edge learning/intelligence\cite{Park19,KBL22,Feng18,Feng19} emerge, in which dedicated fusion centers (FCs), located possibly at edge servers or base stations (BSs), are employed to swiftly aggregate the massive data distributed in WDs, and accordingly make inference about the physical environments for facilitating further actions. In this case, how to compute functional values based on the distributed data from different WDs is becoming a challenging task faced by the FC. For instance, in distributed sensing \cite{Cui07,Dey11}, the FC is interested in retrieving parameters based on the sensing data of multiple geographically distributed WDs. In distributed edge learning (particularly the federated edge learning)\cite{Zhu20-Mag,Zhou-ProcIEEE19}, the edge server coordinates multiple WDs to train shared machine learning models iteratively, and in each iteration, the edge server needs to update the global machine learning model parameters by computing the (weighted) mean values of the local gradients or model parameters at the WDs (or edge devices).

 Conventionally, such distributed functional computation is implemented in wireless networks based on a separate communication and computation design principle, in which different WDs send their data to the FC individually, and the FC first decodes their individual data, and then performs the functional computation. Due to the harmful inter-user interference in multiuser communication, the separate design, however, may incur huge communication overhead and severe transmission delay, especially when the number of WDs becomes large. Recently, {\em over-the-air computation} (AirComp) \cite{Nazer07,Aba16,Gold13-1,Gold13,Gold15} has emerged as a promising solution to compute various functions (especially the so-called {\em nomographic} functions\cite{Gold15}) over the air, by exploiting the signal superposition property of wireless multiple access channels. In AirComp, different WDs can simultaneously transmit their data to the FC over the same frequency band. With proper pre-processing and phase/power control at WDs, the FC can then reliably reconstruct/estimate the desired nomographic function values from the superimposed signals\cite{Gold15}. Different from the conventional separate communication and computation design by combating against the harmful inter-user interference, the new AirComp solution integrates the wireless communication and functional computation into a joint design by utilizing the inter-user interference beneficially, thereby achieving higher resource utilization efficiency.

 In general, there are two different types of AirComp approaches in the literature, namely the uncoded (analog)\cite{Gastpar08} and coded AirComp\cite{Vish12,Wagner08}, respectively. In particular, under independent and identically distributed (i.i.d.) Gaussian sources and a standard Gaussian multiple access channel, the analog AirComp was shown to be optimal in minimizing the computation mean squared error (MSE) distortion\cite{Gastpar08}. Under other setups with, e.g., correlated Gaussian sources\cite{Vish12,Wagner08}, the coded AirComp with sophisticated joint source and channel coding is generally required for minimizing the average MSE distortion. Although sub-optimal in general, analog AirComp has been widely studied for sensing-data related functional computation in internet-of-things (IoT) networks\cite{UAV-AirComp21,Shi20-1,Chen18,Zhu19,Liu20,Cao19,Chen18-2,Zhai20} due to its simplicity in implementation. The (analog) AirComp technique has also attracted growing attentions in emerging distributed (federated) edge learning systems \cite{Shi20,Gunduz20-2,KK-Wong21,Zhu20,Tao20,Zhou21,Cao22,Cao21}, in order to enhance the communication efficiency for frequent gradient/model aggregation. Despite such research progresses, however, these prior works \cite{Chen18,Shi20-1,Zhu19,Liu20,Cao19,Chen18-2,Zhai20,Shi20,Gunduz20-2,KK-Wong21,Zhu20,Tao20,Zhou21,Cao22,Cao21,UAV-AirComp21} mainly focused on the AirComp designs in a single-cell multiple access channel over a small area.


 In practice, to fully exploit the big data value, it is of great importance to aggregate massive data distributed in WDs over a large area. In this scenario, however, the conventional single-cell AirComp design with direct transmission from the WDs to the FC may not work well, as the WDs that are far apart from the FC may suffer from severe signal propagation loss, thus significantly exacerbating the computational distortion\cite{Zhu21}. In the literature, there have been a handful of prior works investigating the large-scale AirComp by considering relays\cite{Zhou20}, multi-cell transmission\cite{Cao20}, and device-to-device (D2D) communications \cite{Feng21,Gunduz20-1,Gunduz20,Shi21-ISIT,Xing21-JASC}, respectively. For instance, the authors in\cite{Zhou20} studied the relay selection in a relay-aided AirComp system, in which one single relay is chosen to amplify and forward its received signal to the FC. The work~\cite{Cao20} studied a multi-cell AirComp system with each BS computing a different function, in which the Pareto boundary for the achievable computation-MSE region at different cells is characterized, and both centralized and distributed power control schemes are developed to achieve Pareto-optimal MSE tuples by balancing the intra-cell signal alignment error and the inter-cell interference. Furthermore, the works \cite{Gunduz20-1,Gunduz20,Shi21-ISIT,Xing21-JASC} investigated the AirComp for federated learning over D2D networks, where analog D2D transmissions are considered for over-the-air gradient averaging and consensus. In addition, under a decentralized MapReduce framework, an AirComp-based transceiver design was developed for computing large-scale nomographic functions over a multi-hop D2D network\cite{Feng21}.

\begin{figure}
  \centering
  \includegraphics[width = 5.5in]{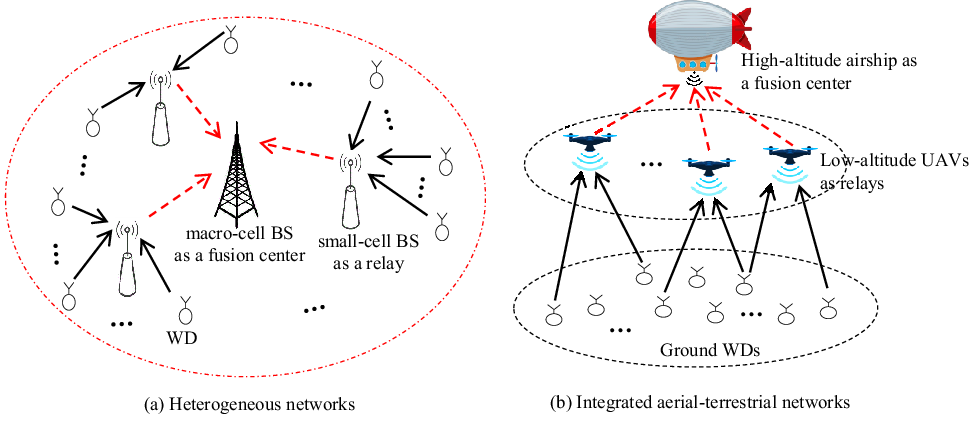}
 \vspace{-0.5cm}
 \caption{Illustrations of hierarchical AirComp systems with two-phase amplify-and-forward (AF) relaying.} \label{fig.Sys-model}
 \end{figure}

 Different from prior works on relay-aided or multi-cell/D2D AirComp systems, this paper presents a new {\em hierarchical} AirComp architecture over a large area, in which {\em multiple} intermediate relays are exploited to help the FC swiftly aggregate data from a large number of WDs for over-the-air functional computation. For example, the hierarchical AirComp can be practically implemented in heterogeneous networks and integrated aerial-terrestrial networks as shown in Figs.~\ref{fig.Sys-model}(a) and \ref{fig.Sys-model}(b), in which the small-cell base stations (BS) and low-altitude unmanned aerial vehicles (UAVs) act as relays, and the macro-cell BS and high-altitude platform station or airship\footnote{See, e.g., the Softbank HAPSMobile at \url{https://www.hapsmobile.com/en/}, the Alibaba Cloud IoT in the Sky LoRa Station at \url{https://www.alibabacloud.com/blog/alibaba-cloud-speeds-up-iot-strategy-at-the-computing-conference_594071}, and GSMA's whitepaper at \url{https://www.gsma.com/futurenetworks/resources/high-altitude-platform-systems-haps-whitepaper-2021/}.}  act as the FC to collect data, respectively.

 In particular, we focus on the hierarchical AirComp system consisting of multiple WDs, multiple relays, and one single FC, by considering a two-phase amplify-and-forward (AF) relaying protocol. In the first phase, the WDs simultaneously broadcast their data to the relays, and in the second phase, the relays amplify the received signals and concurrently forward them to the FC for aggregation. Under this setup, we aim to minimize the computational MSE of the aggregated signal at the FC, subject to the individual power constraints at the WDs and relays, by jointly optimizing the WD transmit coefficients, the relay AF coefficients, and the FC de-noising factor. The main results of this work are summarized as follows.
\begin{itemize}
\item First, we consider the centralized design with global channel state information (CSI), in which the FC perfectly knows the global CSI to coordinate the system design in a centralized manner. Accordingly, the inter-WD interference cross different relays can be beneficially harnessed as useful signals for reducing the computational MSE at the FC. We develop a centralized algorithm to solve the resultant non-convex MSE minimization problem based on alternating optimization, where the optimization of the WD transmit coefficients and the relay AF coefficients (together with the FC de-noising factor) are alternately implemented until convergence. The optimized solution shows that the transmit phase of each WD is opposite to the composite WD-relay-FC channel phase, such that they can be aligned at the FC. It is also shown that the transmit power of a WD or relay follows a regularized composite-channel-inversion structure to strike a balance between minimizing the signal-magnitude-misalignment-induced error and the noise-induced error, where at least one relay's power constraint needs to be activated.

\item Next, in order to alleviate the signaling overhead of obtaining the global CSI and enhance the scalability, we consider a decentralized design with partial CSI, in which the FC and relays make their decisions based on the channel power gain information among different relays. Due to the lack of inter-relay channel phase information in this case, the FC and relays have to treat the inter-relay signal as undesired noise. We propose a decentralized algorithm to efficiently solve the corresponding computational MSE minimization problem, which is implemented in an iterative manner. In each iteration, different relays take turns to optimize the transmit coefficients of their respectively associated WDs without inter-relay coordination, and then the FC optimizes its de-noising factor and the relay AF coefficients. The proposed decentralized algorithm is guaranteed to converge, as the computational MSE is shown to be monotonically non-increasing over each iteration.

\item Finally, numerical results are provided to show the effectiveness of the proposed centralized and decentralized hierarchical AirComp designs. It is shown that both the centralized and decentralized designs achieve significant MSE performance gains over benchmark schemes without the joint optimization. It is also shown that the decentralized design (combating against inter-relay interference) is beneficial in significantly reducing the signaling overhead as composed to the centralized design (harnessing inter-relay interference), at the cost of slightly compromised MSE performance.
\end{itemize}

 The remainder of the paper is organized as follows. Section II introduces the hierarchial AirComp system model and formulates the computational MSE minimization problems with global and partial CSI, respectively. Section III presents the centralized alternating-optimization-based solution in the case with global CSI. Section IV presents the decentralized solution in the case with partial CSI. Section~V provides numerical results to demonstrate the effectiveness and merits of the proposed designs, followed by the conclusion in Section VI.

{\em Notations}: Boldface upper-case letters denote matrices, and boldface lower-case letters denote column vectors. $\mathbb{C}^{m\times n}$ denotes the set of $m\times n$ matrices with complex-valued entries. The superscripts $T$ and $H$ denote the transpose and conjugate transpose operations, respectively. For a scalar $x$, $|x|$ and $\angle{x}$ denote its absolute value and polar angle, respectively. ${\cal CN}(\mu,\sigma^2)$ denotes the distribution of a circular symmetric complex Gaussian (CSCG) random variable with mean $\mu$ and variance $\sigma^2$. ${\cal U}(a,b)$ denotes the uniform distribution within interval $[a,b]$. $\sim$ stands for ``distributed as''. $j=\sqrt{-1}$ represents the imaginary unit. ${\rm diag}(x_1,\ldots,x_n)$ stands for a diagonal matrix whose diagonal entries are $x_1,\ldots,x_n$.

\section{System Model and Problem Formulation}

 As shown in Fig. \ref{fig.Sys-model}, we consider a hierarchical AirComp network over a large area, in which the FC is interested in aggregating the data distributed in $K$ WDs assisted by $M$ intermediate relays. For exposition, all the nodes in the system are assumed to be equipped with a single antenna. Let ${\cal K}\triangleq \{1,\ldots,K\}$ and ${\cal M}\triangleq \{1,\ldots,M\}$ denote the sets of WDs and relays, respectively. It is assumed that each WD $k\in{\cal K}$ collects certain raw data (e.g., temperature, pressure, humidity, viscosity, motion) and needs to transmit them to the FC through the $M$ relays via AirComp. For notational convenience, suppose that each WD $k\in{\cal K}$ is associated with one relay. Denote by ${\cal K}_m\subseteq{\cal K}$ the set of WDs associated with relay $m\in{\cal M}$. We have $\bigcup_{m\in{\cal M}}{\cal K}_m={\cal K}$ and ${\cal K}_{m^\prime}\cap{\cal K}_m=\emptyset$, $\forall m^\prime \neq m$, $m,m^\prime\in{\cal M}$. Let $x_{k}\in\mathbb{C}$ denote WD $k$'s sensing data. Here, the sensing data $\{x_{k}\}_{k\in{\cal K}_m,m\in{\cal M}}$ of the $K$ WDs are assumed to be independent random variables with zero-mean and variances $\{\delta_k^2\}_{k\in{\cal K}_m,m\in{\cal M}}$, i.e., $\mathbb{E}\{|x_{k}|^2\}=\delta_k^2$ and $\delta_k>0$, $\forall k\in{\cal K}_m,m\in{\cal M}$. The FC aims to compute the arithmetic average value\footnote{Note that although only the arithmetic average function is considered in this paper, the proposed designs are also applicable for other nomographic functions such as weighted sum, multiplication, and geometric average\cite{Nazer07}.} of the $K$ WDs' sensing data, i.e., $\bar{x}= \frac{1}{K}\sum_{m\in{\cal M}}\sum_{k\in{\cal K}_m} x_{k}$.

 We consider a quasi-static wireless channel model, in which the wireless channels remain unchanged during the transmission block of our interest. Under this setup, we consider a two-phase AF relaying protocol, in which the transmission block is divided into two phases with equal duration. The transmission over the two phases are detailed in the following.

 In the first phase, the $K$ WDs transmit their signals to the $M$ relays simultaneously. Let $\alpha_{k}\in\mathbb{C}$ and $P_k$ denote the complex-valued transmit coefficient and the maximum power budget of WD $k$, respectively. Accordingly, we have the individual power constraints of the $K$ WDs as
 \begin{align}\label{eq.device-power-constraint}
  \mathbb{E}_{x_k}\left\{|\alpha_k x_k|^2\right\} = |\alpha_{k}|^2 \delta_k^2\leq P_{k},~\forall k\in{\cal K}_m,m\in{\cal M},
\end{align}
 where the expectation in \eqref{eq.device-power-constraint} is taken over the randomness of the sensing data $x_k$. Let $h_{m,k}\in\mathbb{C}$ denote the channel coefficient from WD $k$ to relay $m$. The received signal of relay $m$ is then expressed as
\begin{align}\label{eq.rm}
r_{m} = \underbrace{\sum_{k\in{\cal K}_m} h_{m,k}\alpha_{k} x_{k}}_{\text{intra-relay~signal}} + \underbrace{\sum_{m^\prime\in{\cal M}\setminus\{m\}} \sum_{i\in{\cal K}_{m^\prime}}h_{m,i}\alpha_{i} x_{i}}_{\text{inter-relay~signal}} + z_m,~\forall m\in{\cal M},
\end{align}
 where $z_{m}\sim{\cal CN}(0,\sigma_{m}^2)$ denotes the additive white Gaussian noise (AWGN) at the receiver of relay $m\in{\cal M}$.

 In the second phase, the $M$ relays amplify the respectively received signals $\{r_{m}\}_{m\in{\cal M}}$ and forward them concurrently to the FC. Let $\beta_m\in\mathbb{C}$ denote the complex-valued AF coefficient at each relay $m\in{\cal M}$. Then, the transmit signal of relay $m$ at the second phase is
\begin{align}\label{eq.x_R}
 x_{R,m} =\beta_m r_{m},~\forall m\in{\cal M}.
\end{align}
 Let $P_{R,m}$ denote the maximum transmit power budget at relay $m$. The individual power constraints at the $M$ relays are
\begin{align}\label{eq.relay-power-constraint}
|\beta_m|^2\Big( \sum_{m^\prime\in{\cal M}}\sum_{k\in{\cal K}_{m^\prime}}|\alpha_{k}|^2|h_{m,k}|^2\delta_k^2 + \sigma_{m}^2\Big) \leq P_{R,m},~\forall m\in{\cal M}.
\end{align}
 Furthermore, let $g_m\in\mathbb{C}$ denote the channel coefficient from relay $m$ to the FC. The received signal at the FC is given by
\begin{subequations}\label{eq.y}
\begin{align}
&y=\sum_{m\in{\cal M}} g_mx_{R,m}+z_0 \\
&=\sum_{m\in{\cal M}} \sum_{k\in{\cal K}_m}\alpha_{k}\beta_m g_m h_{m,k}x_{k} + \sum_{m\in{\cal M}} \sum_{m^\prime\in{\cal M}\setminus\{m\}}\sum_{i\in{\cal K}_{m^\prime}}\alpha_{i}\beta_m g_m h_{m,i}x_{i} + \sum_{m\in{\cal M}} \beta_m g_m z_{m} + z_0,
\end{align}
\end{subequations}
 where (\ref{eq.y}b) is obtained by substituting \eqref{eq.x_R} into (\ref{eq.y}a), and $z_0 \sim {\cal CN}(0,\sigma_0^2)$ denotes the AWGN at the FC receiver.

 Upon receiving $y$ in \eqref{eq.y}, the FC starts to reconstruct the target functional value $\bar{x}$. Specifically, the received signal $y$ is directly divided by a positive real-valued {\em de-noising factor}, denoted by $\eta>0$. Accordingly, the estimation of the targeted arithmetic average value $\bar{x}$ at the FC is expressed as
 \begin{align}\label{eq.x-hat}
 \hat{x} = \frac{1}{K\eta} y.
 \end{align}

 Notice that the design of the transmit coefficients $\{\alpha_{k}\}_{k\in{\cal K}_{m},m\in{\cal M}}$ of the $K$ WDs, the AF coefficients $\{\beta_{m}\}_{m\in{\cal M}}$ of the $M$ relays, and the de-noising factor $\eta$ of the FC highly depends on the availability of the CSI $(\{h_{m,k}\}_{k\in{\cal K}_m,m\in{\cal M}},\{g_m\}_{m\in{\cal M}})$. In particular, we focus on the hierarchical AirComp designs by considering two scenarios with global and partial CSI, respectively.

\subsection{Centralized Design with Global CSI}
 First, we consider the centralized design with global CSI, where the CSI $\{h_{m,k}\}_{m\in{\cal M}}$ from each WD $k\in{\cal K}_m$ to the $M$ relays and the CSI $\{g_m\}_{m\in{\cal M}}$ from the relays to the fusion are all perfectly available at a central node (i.e., the FC). Accordingly, the central node can coordinate the transmission of the WDs and relays in a centralized manner, such that the inter-relay signals at the lower layer (i.e., the inter-relay signal in \eqref{eq.rm}) can be exploited beneficially to facilitate the functional computation at the FC.

 Accordingly, the computational MSE to measure the distortion performance between the FC's reconstructed $\hat{x}$ in \eqref{eq.x-hat} and the ground truth $\bar{x}$ is expressed as
\begin{align}\label{eq.mse0}
{\tt MSE}^{\text{I}}(\bm \alpha^{\text{I}},\bm \beta^{\text{I}},\eta^{\text{I}}) &\triangleq \mathbb{E}_{\{x_k,z_m,z_0\}}\left\{|\hat{x}-\bar{x}|^2\right\}  \notag \\
&=\frac{1}{K^2}\Bigg[ \underbrace{
\sum_{m\in{\cal M}}\sum_{k\in{\cal K}_m} \Big|\frac{\alpha^{\text{I}}_{k}\bm h_k^T\bm\Lambda_g{\bm \beta^{\text{I}}}}{\eta^{\text{I}}}-1\Big|^2
\delta_k^2}_{\text{signal-magnitude-misalignment-induced~error}} + \underbrace{\frac{\bm (\beta^{\text{I}})^H \bm\Lambda^H_g\bm\Lambda_{\sigma^2}\bm\Lambda_g\bm \beta^{\text{I}}+\sigma_0^2}{(\eta^{\text{I}})^2}}_{\text{noise-induced~error}}\Bigg],
\end{align}
 where $\bm \alpha^{\text{I}} \triangleq [\alpha^{\text{I}}_1,...,\alpha^{\text{I}}_K]^T$, $\bm \beta^{\text{I}}\triangleq [\beta^{\text{I}}_1,\ldots,\beta^{\text{I}}_M]^T$, $\bm\Lambda_g \triangleq \text{diag}(g_1,\ldots,g_M)$, $\bm\Lambda_{\sigma^2}\triangleq \text{diag}(\sigma^2_{1},\ldots,\sigma^2_{M})$, ${\bm h}_{k}\triangleq [h_{1,k},\ldots,h_{M,k}]^T$, $\forall k\in{\cal K}$, and the expectation is taken on the randomness of all the $K$ WDs' data $\{x_{k}\}_{k\in{\cal K}_m,m\in{\cal M}}$ and the noise $(\{z_m\}_{m\in{\cal M}},z_0)$. It is observed in \eqref{eq.mse0} that ${\tt MSE}^{\text{I}}(\bm \alpha^{\text{I}},\bm \beta^{\text{I}},\eta^{\text{I}})$ consists of the signal-magnitude-misalignment-induced error (the first term) and the noise-induced error (the second term).

 In the centralized design with global CSI, we aim to minimize ${\tt MSE}^{\text{I}}(\bm \alpha^{\text{I}},\bm \beta^{\text{I}},\eta^{\text{I}})$ in \eqref{eq.mse0}, by jointly optimizing the transmit coefficients $\{\alpha_{k}\}_{k\in{\cal K}_m,m\in{\cal M}}$ of the $K$ WDs, the AF coefficients $\{\beta_m\}_{m\in{\cal M}}$ of the $M$ relays, and the de-noising factor $\eta$ of the FC, subject to the individual power constraints in \eqref{eq.device-power-constraint} {\rm and} \eqref{eq.relay-power-constraint} at the WDs and relays, respectively. Mathematically, by omitting the constant factor $1/K^2$ in \eqref{eq.mse0}, the computational MSE minimization problem is formulated as
\begin{subequations}\label{eq.prob-p1}
\begin{align}
({\rm P1}):~
&\min_{\bm \alpha^{\text{I}},\bm\beta^{\text{I}},\eta^{\text{I}}}~
\sum_{m\in{\cal M}}\sum_{k\in{\cal K}_{m}} \Big|\frac{\alpha^{\text{I}}_{k}\bm h_k^T\bm\Lambda_g{\bm \beta^{\text{I}}}}{\eta^{\text{I}}}-1\Big|^2\delta_k^2  + \frac{(\bm \beta^{\text{I}})^H\bm\Lambda^H_g\bm\Lambda_{\sigma^2}\bm\Lambda_g\bm \beta^{\text{I}}+\sigma_0^2}{(\eta^{\text{I}})^2}\\
&~~{\rm s.t.}~~~\eta^{\text{I}}>0~\text{and}~|\alpha^{\text{I}}_{k}|^2 \delta_k^2\leq P_{k},~\forall k\in{\cal K}_m,m\in{\cal M}\\
&~~~~~~~~~|\beta^{\text{I}}_m|^2\Big( \sum_{m^\prime\in{\cal M}}\sum_{k\in{\cal K}_{m^\prime}}|\alpha^{\text{I}}_{k}|^2|h_{m,k}|^2\delta_k^2 + \sigma_{m}^2\Big) \leq P_{R,m},\forall m\in{\cal M}.
\end{align}
\end{subequations}
 Due to the coupling of $\bm \alpha^{\text{I}}$ and $\bm \beta^{\text{I}}$ in the objective function (\ref{eq.prob-p1}a) and constraints (\ref{eq.prob-p1}c), problem (P1) is a non-convex optimization problem, thus making the globally optimal solution difficult to obtain. As such, we employ a centralized alternating-optimization-based approach to obtain a high-quality solution to problem (P1), as will be shown in Section~III.

\subsection{Decentralized Design with Partial CSI}
 Next, to avoid the signaling overhead in obtaining global CSI, we consider an alternative design with partial CSI. In particular, we consider that each relay $m\in{\cal M}$ only knows the CSI $g_m$ from itself to the FC and $\{h_{k,m}\}_{k\in{\cal K}_m}$ of its associated WDs in set ${\cal K}_m$, as well as the magnitude knowledge $\{|h_{m^{\prime},k}|\}_{m^\prime\in{\cal M}\setminus\{m\}}$ from each WD $k\in{\cal K}_m$ to the $(M-1)$ non-associated relays, and the magnitude knowledge $\{|g_{m^\prime}|\}_{m^\prime\in{\cal M}\setminus\{m\}}$ from the other $(M-1)$ relays to the FC. We also consider that the FC only knows the magnitude knowledge $\{|h_{k,m}|\}_{k\in{\cal K},m\in{\cal M}}$ and $\{|g_m|\}_{m\in{\cal M}}$. Notice that the channel magnitude changes over time at a much slower scale than the channel phase in general\cite{Goldsmith05}. Therefore, as compared to obtaining the global CSI, the $M$ relays and the FC can obtain the channel magnitude information based on long-term channel measurements, thus significantly reducing signalling exchange overhead. In addition, due to the fast changing nature of the channel phase, when the channel phase information is not available, we assume that the phase of channel coefficient $h_{m^\prime,k}$ follows a uniform distribution within interval $[0,2\pi]$, i.e., $\angle{h_{m^\prime,k}}\sim{\cal U}(0,2\pi)$, $\forall k\in{\cal K}_m$, $m^\prime\neq m\in{\cal M}$.

 Due to the lack of phase information of the inter-relay links, the corresponding inter-relay signals (i.e., $\{\sum_{m^\prime\in{\cal M}\setminus\{m\}} \sum_{i\in{\cal K}_{m^\prime}}h_{m,i}\alpha_{i} x_{i}\}_{m\in{\cal M}}$) in \eqref{eq.rm}  become harmful interference in this case. This is quite different from the centralized design with global CSI, where the inter-relay signals can be fully exploited in reduce the computational MSE. Accordingly, with partial CSI, the computational MSE between the FC's reconstructed function value $\hat{x}$ and the ground truth $\bar{x}$ is expressed as
\begin{align}\label{eq.mse-new1}
 &{\tt MSE}^{\text{II}}(\bm \alpha^{\text{II}},\bm \beta^{\text{II}},\eta^{\text{II}})  \triangleq \mathbb{E}_{\{x_k,z_m,z_0,\angle{h_{m^{\prime},k}} \}}\Big\{\Big|\hat{x} -\bar{x}\Big|^2\Big\} \notag\\
 &\quad= \frac{1}{K^2}\Bigg[
\sum_{m\in{\cal M}}\sum_{k\in{\cal K}_m} \Big|
\frac{\alpha^{\text{II}}_{k}\beta^{\text{II}} _m g_m h_{m,k}}{\eta^{\text{II}}} -1 \Big|^2\delta_{k}^2 + \underbrace{\sum_{m\in{\cal M}} \sum_{m^\prime\in{\cal M}\setminus\{m\}}\sum_{i\in{\cal K}_{m^\prime}}\frac{|\alpha^{\text{II}} _i|^2|\beta^{\text{II}}_m|^2 |g_m|^2 |h_{m,i}|^2\delta_i^2}{(\eta^{\text{II}}) ^2} }_{\substack{\text{inter-relay-interference-induced error}}} \notag \\
&\quad~~~~~~~~ + \frac{\sum_{m\in{\cal M}}|\beta^{\text{II}} _m|^2|g_m|^2\sigma_m^2+\sigma_0^2}{(\eta^{\text{II}})^2}\Bigg],
 \end{align}
 where $\bm \alpha^{\text{II}}\triangleq [\alpha_1^{\text{II}},...,\alpha_K^{\text{II}}]^T$, $\bm\beta^{\text{II}}\triangleq [\beta^{\text{II}}_1,...,\beta^{\text{II}}_M]^T$, and the expectation is taken over the randomness of the $K$ WDs' sensing data $\{x_{k}\}_{k\in{\cal K}_m,m\in{\cal M}}$, the phases of the inter-relay CSI $\{h_{m^{\prime},k}\}_{k\in{\cal K}_{m},m^\prime\neq m}$, and the noise $(\{z_m\}_{m\in{\cal M}},z_0)$. Compared with ${\tt MSE}^{\text{I}}(\bm \alpha^{\text{I}},\bm \beta^{\text{I}},\eta^{\text{I}})$ in \eqref{eq.mse0} only including the signal-magnitude-misalignment-induced error and the noise-induced error, ${\tt MSE}^{\text{II}}(\bm \alpha^{\text{II}},\bm \beta^{\text{II}},\eta^{\text{II}})$ in \eqref{eq.mse-new1} additionally includes the inter-relay-interference-induced error.

 By omitting the constant factor $\frac{1}{K^2}$ in ${\tt MSE}^{\text{II}}(\bm \alpha^{\text{II}},\bm \beta^{\text{II}},\eta^{\text{II}})$, the computational MSE minimization problem with partial CSI is formulated as
 \begin{subequations}\label{eq.prob-p2-new}
 \begin{align}
&({\rm P2}): \notag \\
&\min_{\bm \alpha^{\text{II}},\bm\beta^{\text{II}},\eta^{\text{II}}}\sum_{m\in{\cal M}} \sum_{k\in{\cal K}_m}\Big|
 \frac{\alpha^{\text{II}}_{k}\beta^{\text{II}}_m g_m h_{m,k}}{\eta^{\text{II}}}  -1\Big|^2\delta_{k}^2 + \sum_{m\in{\cal M}} \sum_{m^\prime\in{\cal M}\setminus\{m\}}\sum_{i\in{\cal K}_{m^\prime}}\frac{|\alpha^{\text{II}} _i|^2|\beta^{\text{II}} _m|^2 |g_m|^2 |h_{m,i}|^2|\delta_i|^2}{(\eta^{\text{II}}) ^2}  \notag \\
 &\quad\quad\quad\quad\quad\quad\quad +  \frac{\sum_{m\in{\cal M}}|\beta^{\text{II}}_m|^2|g_m|^2\sigma_m^2 + \sigma_0^2}{(\eta^{\text{II}})^2}  \\
&~ {\rm s. t.}~\eta^{\text{II}}>0~\text{and}~|\alpha^{\text{II}}_{k}|^2\delta_{k}^2\leq P_{k},~\forall k\in{\cal K}_m,m\in{\cal M}\\
&\quad |\beta^{\text{II}}_m|^2 \Big( \sum_{k\in{\cal K}_m}|\alpha^{\text{II}}_{k}|^2|h_{m,k}|^2\delta_{k}^2 +\sum_{m^\prime\in{\cal M}\setminus\{m\}} \sum_{i\in{\cal K}_{m^\prime}} |\alpha^{\text{II}} _i|^2|h_{m,i}|^2\delta_i^2 + \sigma_{m}^2 \Big) \leq P_{R,m},\forall m\in{\cal M}.
 \end{align}
 \end{subequations}
 Similarly to problem (P1), problem (P2) is a non-convex optimization problem due to the coupling of $\bm \alpha^{\text{II}}$ and $\bm \beta^{\text{II}}$ in the objective function (\ref{eq.prob-p2-new}a) and constraints (\ref{eq.prob-p2-new}c). We will develop a decentralized algorithm for efficiently solving problem (P2) with partial CSI in Section~IV.

\section{Centralized Design Solution to (P1) with Global CSI}
 In this section, we present a centralized algorithm to solve problem (P1) with global CSI, by alternately optimizing $\bm \alpha^{\text{I}}$ and $(\bm \beta^{\text{I}},\eta^{\text{I}})$ in an iteration manner.

\subsection{Optimization of WD Transmit Coefficients $\bm \alpha^{\text{I}}$ with Given $(\bm \beta^{\text{I}},\eta^{\text{I}})$}
 Under given AF coefficient vector $\bm\beta^{\text{I}}$ of the $M$ relays and the de-noising factor $\eta^{\text{I}}$ of the FC, we optimize the transmit coefficient vector $\bm \alpha^{\text{I}}$ of the $K$ WDs. Notice that the noise-induced error $((\bm \beta^{{\text{I}}})^H\bm\Lambda^H_g\bm\Lambda_{\sigma^2}\bm\Lambda_g\bm \beta^{\text{I}}+\sigma_0^2)/(\eta^{\text{I}})^2$ in \eqref{eq.mse0} is independent of the WD transmit coefficients $\{\alpha^{\text{I}}_{k}\}_{k\in{\cal K}_m,m\in{\cal M}}$. By ignoring this term, the optimization of $\bm \alpha^{\text{I}}$ is equivalently expressed as
\begin{subequations}\label{eq.prob-p2}
\begin{align}
&\min_{\bm \alpha^{\text{I}}}~\sum_{m\in{\cal M}}\sum_{k\in{\cal K}_m} \left|\frac{\alpha^{\text{I}}_{k}\bm h_k^T\bm\Lambda_g{\bm \beta^{\text{I}}}}{\eta^{\text{I}}}-1\right|^2\delta_k^2 \\
&~{\rm s.t.}~~ |\alpha^{\text{I}}_{k}|^2\delta_k^2\leq P_{k},~\forall k\in{\cal K}\\
&~~~~~~\sum_{m\in{\cal M}}\sum_{k\in{\cal K}_m} |\alpha^{\text{I}}_{k}|^2|h_{m,k}|^2\delta_k^2 + \sigma_{m}^2
\leq \frac{P_{R,m}}{|\beta_m|^2},~\forall m\in{\cal M}.
\end{align}
\end{subequations}
 Let $\bm \alpha^{\text{I*}}\triangleq [\alpha_1^{\text{I*}},...,\alpha_K^{\text{I*}}]^T$ denote the optimal solution to problem \eqref{eq.prob-p2}.

 To start with, we establish the following lemma to characterize the optimal phases of $\{\alpha_{k}^{\text{I*}}\}$.

\begin{lemma}\label{prop.alpha}
At the optimal solution to problem \eqref{eq.prob-p2}, it must hold that
\begin{align}\label{eq.alpha_phase_I}
\angle\alpha^{\text{I*}}_{k} = -\angle{\bm h_k^T\bm\Lambda_g{\bm \beta^{\text{I}}}},~\forall k\in{\cal K}_m,m\in{\cal M}.
\end{align}
\end{lemma}

 \begin{IEEEproof}
 See Appendix~\ref{prop.alpha-proof}.
 \end{IEEEproof}

 Lemma~\ref{prop.alpha} indicates that, in order to compute the arithmetic average function of $\{x_k\}_{k\in{\cal K}_m,m\in{\cal M}}$ distributed at the distributed WDs over the air, each WD $k\in{\cal K}$ should adjust the phase of transmit coefficient $\alpha^{\text{I}}_k$ to be opposite to that of the composite WD-relay-FC channel (i.e., $\sum_{m\in{\cal M}} h_{m,k}g_m\beta^{\text{I}}_m$), such that all the $K$ WDs' signal phases can be aligned at the FC, thus leading to a {\em constructive} addition to facilitate the functional computation.

 Based on Lemma~\ref{prop.alpha}, we define $\alpha^{\text{I}}_k \triangleq \bar{\alpha}^{\text{I}}_k e^{j\angle{\alpha^{\text{I*}}_{k}}}$, $\forall k\in{\cal K}_m,m\in{\cal M}$, where $\bar{\alpha}^{\text{I}}_k$ denotes the magnitude of WD $k$'s transmit coefficient, i.e., $\bar{\alpha}^{\text{I}}_k=|{\alpha}^{\text{I}}_k|$. By substituting $\{\alpha^{\text{I}}_k =\bar{\alpha}^{\text{I}}_k e^{j\angle{\alpha^{\text{I*}}_{k}}}\}_{k\in{\cal K}_m, m\in{\cal M}}$, problem~\eqref{eq.prob-p2} is recast as
 \begin{subequations}\label{eq.prob-p21}
 \begin{align}
 &\min_{ \bar{\bm \alpha}^{\text{I}}}~~\sum_{m\in{\cal M}}\sum_{k\in{\cal K}_m} \Big(\bar{\alpha}^{\text{I}}_{k}\frac{|\bm h_k^T\bm\Lambda_g{\bm \beta^{\text{I}}}|}{\eta^{\text{I}}}-1\Big)^2\delta_k^2 \\
 &~{\rm s.t.}~ 0\leq \bar{\alpha}^{\text{I}}_{k} \leq \sqrt{P_{k}/\delta^2_k},~~\forall k\in{\cal K}\\
 &~~~~~\sum_{m\in{\cal M}}\sum_{k\in{\cal K}_m} (\bar{\alpha}^{\text{I}}_{k})^2|h_{m,k}|^2\delta_k^2 + \sigma_{m}^2
\leq \frac{P_{R,m}}{|\beta^{\text{I}}_m|^2},~\forall m\in{\cal M},
 \end{align}
 \end{subequations}
 where $\bar{\bm \alpha}^{\text{I}}\triangleq [\bar{\alpha}^{\text{I}}_1,...,\bar{\alpha}^{\text{I}}_K]^T$. As problem \eqref{eq.prob-p21} is convex and satisfies the Slater's conditions, then strong duality holds between \eqref{eq.prob-p21} and its dual problem~\cite{Boyd_book}. As a result, one can solve problem~\eqref{eq.prob-p21} by leveraging the Karush-Kuhn-Tucker (KKT) optimality conditions. Let $\bar{\bm \alpha}^{\text{I*}}$ denote the optimal primal solution to problem~\eqref{eq.prob-p21}, and $\bm\mu^*=[\mu_1^*,...,\mu_M^*]^T$ the optimal dual solution associated with the $M$ constraints in (\ref{eq.prob-p21}c). In the following, we establish the proposition for the optimal solution to problem~\eqref{eq.prob-p21}.

\begin{proposition}\label{prop.alpha-value}
The optimal solution $\bar{\bm \alpha}^{\text{I*}}$ to problem~\eqref{eq.prob-p21} is given as
\begin{align}\label{eq.alpha_magnitude}
\bar{\alpha}^{\text{I*}}_{k} & = \min\Bigg\{ \frac{\frac{|\bm h_k^T\bm\Lambda_g\bm \beta^{\text{I}}|}{\eta^{\text{I}}}}{\frac{|\bm h_k^T\bm\Lambda_g\bm \beta^{\text{I}}|^2}{(\eta^{\text{I}})^2} + \sum_{m\in{\cal M}} \mu^{*}_m|h_{m,k}|^2},\sqrt{\frac{P_{k}}{\delta^2_k}}\Bigg\},~\forall k\in{\cal K}_m,m\in{\cal M},
\end{align}
 where $\{\mu^{\text{I*}}_m\}_{m\in{\cal M}}$ are non-negative and satisfy the following complementary slackness conditions:
\begin{align}\label{eq.opt-mu}
\mu^{*}_{m}\Big(\sum_{m^\prime\in{\cal M}}\sum_{k\in{\cal K}_{m^\prime}}(\bar{\alpha}_{k}^{\text{I*}})^2|h_{m,k}|^2\delta^2_k +
\sigma^2_m-\frac{P_{R,m}}{|\beta^{\text{I}}_m|^2}\Big)=0,~\forall m\in{\cal M}.
\end{align}
\end{proposition}
\begin{IEEEproof}
 See Appendix~\ref{prop.alpha-value-proof}.
\end{IEEEproof}

 Propositions~\ref{prop.alpha-value} indicates a regularized channel-inversion structure for the optimal amplitude of transmit coefficient (or equivalently the transmit power) at each WD $k\in{\cal K}_m$, based on the {\em effective} WD-relay-FC channel coefficient $\bm h_k^T\bm\Lambda_g\bm \beta^{\text{I}}$ and the regularization component $\sum_{m\in{\cal M}} \mu^{\rm *}_m|h_{m,k}|^2$ related to the individual power constraints at the relays in (\ref{eq.prob-p21}c). In particular, if both the WDs and relays have sufficiently large power budgets, then it yields that $\mu_m^{\rm *}=0$, $\forall m\in{\cal M}$, and $\bar{\alpha}^{\text{I*}}_{k} = \frac{1}{|\bm h_k^T\bm\Lambda_g\bm \beta^{\text{I}}|/\eta^{\text{I}}}$, $\forall k\in{\cal K}_m,m\in{\cal M}$, i.e., the composite-channel-inversion power control is employed for these WDs.

 By further combining Lemma~\ref{prop.alpha} and Proposition~\ref{prop.alpha-value}, we finally obtain the optimal solution $\bm \alpha^{\text{I*}}$ to problem \eqref{eq.prob-p2} as
\begin{align}\label{eq.alpha-opt}
\alpha^{\text{I*}}_{k}=\bar{\alpha}^{\text{I*}}_{k}e^{-j\angle{\bm h_k^T\bm\Lambda_g\bm \beta^{\text{I}}}},~\forall k\in{\cal K}_m,m\in{\cal M}.
\end{align}

\subsection{Joint Optimization of Relay AF Coefficients $\bm \beta^{\text{I}}$ and FC De-noising Factor $\eta^{\text{I}}$ with Given $\bm\alpha^{\text{I}}$}
 Under given transmit coefficient vector $\bm \alpha^{\text{I}}$ of the $K$ WDs, we jointly optimize the AF coefficient vector $\bm \beta^{\text{I}}$ of the $M$ relays and the de-noising factor $\eta^{\text{I}}$ of the FC. In this case, problem (P1) under the given $\bm \alpha^{\text{I}}$ is reduced into
\begin{subequations}\label{eq.prob-p3}
\begin{align}
&\min_{\bm\beta^{\text{I}},\eta^{\text{I}}}~\sum_{m\in{\cal M}}\sum_{k\in{\cal K}_m} \Big|\frac{\alpha^{\text{I}}_{k}\bm h_k^T\bm \Lambda_g{\bm \beta^{\text{I}}}}{\eta^{\text{I}}}-1\Big|^2\delta_k^2 + \frac{(\bm \beta^{{\text{I}}})^H\bm\Lambda^H_g\bm\Lambda_{\sigma^2}\bm\Lambda_g
\bm \beta^{\text{I}}+\sigma_0^2}{(\eta^{\text{I}})^2}\\
&~~~{\rm s.t.}~~~\eta^{\text{I}}>0~{\text{and}}~|\beta^{\text{I}}_m|^2 \leq \bar{P}_{R,m},~~\forall m\in{\cal M},
\end{align}
\end{subequations}
 where $\bar{P}_{R,m} \triangleq \frac{P_{R,m}}{\sum_{m^\prime\in{\cal M}}\sum_{k\in{\cal K}_{m^\prime}}|\alpha^{\text{I}}_{k}|^2|h_{m,k}|^2\delta_k^2+\sigma_{m}^2}$, $\forall m\in{\cal M}$. Due to the coupling of $\bm \beta^{\text{I}}$ and $\eta^{\text{I}}$ in (\ref{eq.prob-p3}a), problem~\eqref{eq.prob-p3} is non-convex. Nonetheless, with the change of variables, we next show that  problem~\eqref{eq.prob-p3} can be equivalently transformed as a convex optimization problem.

 To this end, we define $\tilde{\bm \beta}^{\text{I}}=[\tilde{\bm \beta}^{\text{I}}_1,...,\tilde{\bm \beta}^{\text{I}}_M]^T \triangleq \frac{\bm \beta^{\text{I}}}{\eta^{\text{I}}}$ and $\gamma^{\text{I}} \triangleq \frac{1}{(\eta^{\text{I}})^2}$. By substituting $\tilde{\bm \beta}^{\text{I}}$ and $\gamma^{\text{I}}$, problem~\eqref{eq.prob-p3} is reformulated as
 \begin{subequations}\label{eq.prob-p3-1}
 \begin{align}
 &\min_{\tilde{\bm \beta}^{\text{I}},\gamma^{\text{I}}}~\sum_{m\in{\cal M}}\sum_{k\in{\cal K}_m} \Big|\alpha^{\text{I}}_{k}\bm h_k^T\bm \Lambda_g \tilde{\bm\beta}^{\text{I}}-1\Big|^2\delta_k^2 + (\tilde{\bm \beta}^{{\text{I}}})^H\bm\Lambda^H_g\bm\Lambda_{\sigma^2}\bm\Lambda_g
 \tilde{\bm \beta}^{\text{I}} + \gamma^{\text{I}}\sigma_0^2\\
 &~~{\rm s.t.}~~|\tilde{\beta}^{\text{I}}_m|^2 \leq \gamma^{\text{I}}\bar{P}_{R,m},~\forall m\in{\cal M}\\
 &\quad\quad~~ \gamma^{\text{I}}>0.
\end{align}
\end{subequations}
Since the objective function in (\ref{eq.prob-p3-1}a) is convex and each constraint in (\ref{eq.prob-p3-1}b) characterizes a convex second-order cone (SOC) with respect to $(\tilde{\bm \beta}^{\text{I}},\gamma^{\text{I}})$ \cite{Boyd_book}, problem \eqref{eq.prob-p3-1} is convex. It is verified that problem \eqref{eq.prob-p3-1} satisfies the Slater's condition\cite{Boyd_book}, i.e., there always exists a feasible solution $(\tilde{\bm \beta}^{\text{I}},\gamma^{\text{I}})$ such that $|\tilde{\beta}^{\text{I}}_m|^2 < \gamma\bar{P}_{R,m}$, $\forall m\in{\cal M}$. Therefore, strong duality holds for problem~\eqref{eq.prob-p3-1}.

 Denote by $(\tilde{\bm \beta}^{\text{I*}},\gamma^{\text{I*}})$ the optimal solution to problem \eqref{eq.prob-p3-1}. Denote by $\bm \lambda^*=[\lambda_1^{*},...,\lambda_M^{*}]^T$ the optimal Lagrange multiplier vector associated with the $M$ constraints in (\ref{eq.prob-p3-1}b). As problem \eqref{eq.prob-p3-1} has a zero duality gap, the KKT optimality conditions are necessary and sufficient for a primal-dual pair $(\tilde{\bm \beta}^{\text{I*}},\gamma^{\text{I*}},\bm\lambda^*)$ to be optimal\cite{Boyd_book}. Based on the KKT optimality conditions, We have the following proposition on the optimal primal-dual pair $(\tilde{\bm \beta}^{\text{I*}},\gamma^{\text{I*}},\bm \lambda^{*})$ for problem \eqref{eq.prob-p3-1}.

 \begin{proposition}\label{Prop.p3-1}
 For problem \eqref{eq.prob-p3-1}, the optimal primal-dual pair $(\tilde{\bm \beta}^{\text{I*}},\gamma^{\text{I*}},\bm\lambda^{*})$ is obtained as
 \begin{subequations}\label{eq.opt-beta-gamma}
 \begin{align}
 &\tilde{\bm \beta}^{\text{I*}} = \bm\Lambda(\bar{\bm H} +\bm \Lambda_{\sigma^2} + \bm\Lambda_{{\lambda}^{*}} )^{-1}\Big(\sum_{m\in{\cal M}}\sum_{k\in{\cal K}_m} \alpha^{\text{I}}_k\bm h_k^T\Big)^H\\
 &\gamma^{\text{I*}} = \frac{|\tilde{\beta}^{\rm *}_{m^{\rm *}}|^2}{\bar{P}_{R,m^{\rm *}}}~~{\rm with}~~m^{\rm *}\triangleq \argmax_{m\in{\cal M}}~
 \frac{|\tilde{\beta}^{\rm *}_m|^2}{\bar{P}_{R,m}}\\
 &\lambda_{m}^{\rm *} = \begin{cases} \frac{\sigma_0^2}{\bar{P}_{R,m}}, &{\rm if}~m = m^{\rm *}\\
 0, &{\rm if}~m \neq m^{\rm *},
 \end{cases}
 \end{align}
 \end{subequations}
 where $\bm\Lambda \triangleq {\rm diag}\big(\frac{g^H_1}{|g_1|^2},...,\frac{g^H_M}{|g_M|^2}\big)$, $\bar{\bm H}\triangleq \sum_{m\in{\cal M}}\sum_{k\in{\cal K}_m}\delta_k^2|\alpha^{\text{I}}_k|^2\bm h_k\bm h_k^H$, and $\bm\Lambda_{{\lambda^{\rm *}}} \triangleq {\rm diag}\big(\frac{\lambda^{\rm *}_1}{|g_1|^2},...,\frac{\lambda^{\rm *}_M}{|g_M|^2}\big)$.
\end{proposition}

 \begin{IEEEproof}
 See Appendix~\ref{Proof-Prop.p3-1}.
 \end{IEEEproof}

 Proposition~\ref{Prop.p3-1} reveals several design insights for achieving the optimal $(\tilde{\bm \beta}^{\text{I*}},\gamma^{\text{I*}})$ to minimize the computational MSE under given transmit coefficients $\{\alpha^{\text{I}}_k\}_{k\in{\cal K}_m,m\in{\cal M}}$ of the $K$ WDs as follows.
 \begin{itemize}
 \item First, the term $(\bar{\bm H} +\bm \Lambda_{\sigma^2} + \bm\Lambda_{{\lambda}^{*}} )^{-1}\big(\sum_{m\in{\cal M}}\sum_{k\in{\cal K}_m} \alpha^{\text{I}}_k\bm h_k^T\big)^H$ in \eqref{eq.opt-beta-gamma} shows that $\tilde{\bm\beta}^{\text{I*}}$ has a minimum MSE (MMSE)-like structure, which balances the tradeoff between maximizing the received signal power from WDs and minimizing the receiver noise power at the relays, and the diagonal matrix $\bm \Lambda$ indicates the relays need to use the channel-inversion structure to counter against the channel fading effect from the relays to the FC.
 \item Second, at the optimality of \eqref{eq.prob-p3-1}, there is only one non-zero dual variable $\lambda_{m^{*}}^{*}=\frac{\sigma_0^2}{\bar{P}_{R,m^{*}}}$ corresponding to relay $m^{*}$ that utilizes its full power for AF relaying. Accordingly, we have $\gamma^{\text{I*}} =\frac{|\tilde{\beta}^{\text{I*}}_{m^{*}}|^2}
     {\bar{P}_{R,m^{*}}}$, which ensures that only one relay uses up its power while the other relays use the channel inversion power control without violating their respective power constraints.
\end{itemize}

 Note that the optimal primal-dual pair $(\tilde{\bm \beta}^{\text{I*}},\gamma^{\text{I*}},\bm\lambda^{*})$ in \eqref{eq.opt-beta-gamma} in Proposition~\ref{Prop.p3-1} is in a recursion from, where a crucial procedure is to determine the optimal relay index $m^{*}$. We can adopt an exhaustive search or bisection search to find the optimal $m^{*}$\cite{Boyd_book}. Once the optimal relay index $m^{*}$ is obtained, then we are ready to compute $(\tilde{\bm \beta}^{\text{I*}},\gamma^{\text{I*}})$ based on (\ref{eq.opt-beta-gamma}a) and (\ref{eq.opt-beta-gamma}b). With $(\tilde{\bm \beta}^{\text{I*}},\gamma^{\text{I*}})$ obtained, the optimal solution $(\bm\beta^{\text{I*}},\eta^{\text{I*}})$ to problem~\eqref{eq.prob-p3} is thus given by
 \begin{subequations} \label{eq.sol-prob-p3}
 \begin{align}
 &\eta^{\text{I*}} = \frac{1}{\sqrt{\gamma^{\text{I*}}}}, \\ 
 &\bm \beta^{\text{I*}} = \eta^{\text{I*}}\tilde{\bm \beta}^{\text{I*}}.
 \end{align}
 \end{subequations}

\subsection{Complete Centralized Algorithm for Solving (P1)}
 By combining Propositions \ref{prop.alpha-value} and \ref{Prop.p3-1} together with \eqref{eq.alpha-opt} and \eqref{eq.sol-prob-p3}, we propose an alternating-optimization-based approach to efficiently obtain a high-quality solution of problem (P1) with global CSI, which is presented as Algorithm~1 in Table~I and is implemented in a centralized manner. For notational convenience, we use ${\tt MSE}^{\text{I}(n)}$ to denote ${\tt MSE}^{\text{I}}(\bm\alpha^{\text{I}(n)},\bm \beta^{\text{I}(n)}, \eta^{\text{I}(n)})$. As problems \eqref{eq.prob-p2} and \eqref{eq.prob-p3} are optimally solved, it is ensured that ${\tt MSE}^{\text{I}(n)}$ is non-increasing over iterations generated by Algorithm~1. Therefore, the convergence of Algorithm~1 is guaranteed. 

\begin{table}[htp]
\begin{center}
\caption{Proposed Centralized Algorithm 1 for Solving Problem (P1) with Global CSI}
\hrule
\begin{itemize}
\item[a)] {\bf Initialization:} $n=1$, $\epsilon>0$, $|\alpha_{k}^{\text{I}(0)}|^2\leq P_{k}$, $\forall k\in{\cal K}_m$, and $\beta^{\text{I}(0)}_m$ $|\beta^{\text{I}(0)}_m|^2\big(\sum_{k=1}^K|\alpha^{\text{I}(0)}_k h_{k,m}|^2+\sigma_{R,m}^2\big) \leq P_{R,m}$, $\forall m\in{\cal M}$.
\item[b)] {\bf While} $({\tt MSE}^{\text{I}(n-1)}-{\tt MSE}^{\text{I}(n)})/{\tt MSE}^{\text{I}(n-1)}>\epsilon$, {\bf do}
 \begin{itemize}
 \item Under given $(\bm \beta^{\text{I}(n-1)},\eta^{\text{I}(n-1)})$, obtain the optimal solution $\{\alpha_{k}^{\text{I}(n)}\}_{k\in{\cal K}_m,m\in{\cal M}}$ for problem~\eqref{eq.prob-p2} based on Proposition~\ref{prop.alpha-value} and ~\eqref{eq.alpha-opt};
\item Under given $\{\alpha_{k}^{\text{I}(n)}\}_{k\in{\cal K}_m,m\in{\cal M}}$, obtain the optimal solution $(\bm \beta^{\text{I}(n)},\eta^{\text{I}(n)})$ for problem~\eqref{eq.prob-p3} based on Proposition~\ref{Prop.p3-1} and~\eqref{eq.sol-prob-p3};
 \item $n=n+1$;
    \end{itemize}
\item[c)] {\bf End while}
\item[e)] {\bf Output:} $(\{\alpha_k^{\text{I}(n)}\}_{k\in{\cal K}_m,m\in{\cal M}},\bm \beta^{\text{I}(n)},\eta^{\text{I}(n)})$ for problem (P1) with global CSI.
\end{itemize}
\hrule
\end{center}
\end{table}

\section{Decentralized Design Solution to (P2) with Partial CSI}
 In this section, we propose an iterative algorithm for solving problem (P2) with partial CSI, which can be implemented in a decentralized fashion among the $M$ relays and the FC.

\subsection{Reformulation of ${\tt MSE}^{\text{II}}(\bm \alpha^{\text{II}},\bm \beta^{\text{II}},\eta^{\text{II}})$ for Problem (P2)}
 First, we reformulate the expression of ${\tt MSE}^{\text{II}}(\bm \alpha^{\text{II}},\bm \beta^{\text{II}},\eta^{\text{II}})$ for problem (P2). By changing the order of algebraic manipulation, it holds that
 \begin{align}\label{eq.order-manipulation}
 \sum_{m\in{\cal M}} \sum_{m^\prime\in{\cal M}\setminus\{m\}}\sum_{i\in{\cal K}_{m^\prime}} |\alpha^{\text{II}} _i|^2 |\beta_m^{\text{II}}|^2 |g_m|^2 |h_{m,i}|^2\delta_i^2  =  \sum_{m\in{\cal M}} \sum_{m^\prime\in{\cal M}\setminus\{m\}} \sum_{k\in{\cal K}_{m}} |\alpha^{\text{II}} _k|^2 |\beta^{\text{II}}_{m^\prime}|^2 |g_{m^\prime}|^2 |h_{m^\prime,k}|^2\delta_k^2.
 \end{align}
 Based on \eqref{eq.order-manipulation}, ${\tt MSE}^{\text{II}}(\bm \alpha^{\text{II}},\bm \beta^{\text{II}},\eta^{\text{II}})$ in \eqref{eq.mse-new1} can be re-written as
 \begin{align}\label{eq.mse-new2}
  &{\tt MSE}^{\text{II}}(\bm \alpha^{\text{II}},\bm \beta^{\text{II}},\eta^{\text{II}}) = \frac{1}{K^2}\Big[ \frac{\sum_{m\in{\cal M}}|\beta^{\text{II}}_m|^2|g_m|^2\sigma_m^2+\sigma_0^2}{(\eta^{\text{II}})^2}\Big] + \frac{1}{K^2}\times \notag \\
  & \sum_{m\in{\cal M}} \Bigg[\sum_{k\in{\cal K}_m}\Big|
 \frac{\alpha^{\text{II}}_{k}\beta^{\text{II}}_m g_m h_{m,k}}{\eta^{\text{II}}}  -1\Big|^2\delta_{k}^2 + \underbrace{\sum_{m^\prime\in{\cal M}\setminus\{m\}} \sum_{k\in{\cal K}_m} \frac{|\alpha^{\text{II}}_k|^2|\beta^{\text{II}}_{m^\prime}|^2|g_{m^\prime}|^2|h_{m^\prime,k}|^2
 \delta_k^2}{(\eta^{\text{II}})^2}}_{\text{inter-relay-interference-induced~error~from relays}~m~{\text{to}}~m^\prime\in{\cal M}\setminus\{m\}} \Bigg],
\end{align}
where the term $\sum_{m^\prime\in{\cal M}\setminus\{m\}} \sum_{k\in{\cal K}_m} \frac{|\alpha^{\text{II}}_k|^2|\beta^{\text{II}}_{m^\prime}|^2|g_{m^\prime}|^2|h_{m^\prime,k}|^2
 \delta_k^2}{(\eta^{\text{II}})^2}$ denotes the inter-relay-interference-induced error from relay $m$ to relay $m^\prime\in{\cal M}\setminus\{m\}$. Building on the reformulated ${\tt MSE}^{\text{II}}(\bm \alpha^{\text{II}},\bm \beta^{\text{II}},\eta^{\text{II}})$ in \eqref{eq.mse-new2}, we present an iterative algorithm for solving problem (P2) with partial CSI, which can be efficiently implemented in a decentralized fashion as follows.

 In each iteration, first, each relay $m\in{\cal M}$ takes turns to optimize the transmit coefficients of its associated WDs, under given AF coefficients of the other relays and de-noising factor of the FC. The information required for relay $m$ to make decision includes $\{h_{m,k}\}_{k\in{\cal K}_m}$, $g_m$, $\{|h_{m^\prime,k}|\}_{m^\prime\in{\cal M}\setminus\{m\},k\in{\cal K}_m}$, and $\{|g_{m^\prime}|\}_{m^\prime\in{\cal M}\setminus\{m\}}$. Correspondingly, each relay $m\in{\cal M}$ needs to solve the following problem:
\begin{subequations}\label{prob-P2m}
 \begin{align}
 &({\rm P3}.m):\notag \\
 &\min_{\{\alpha^{\text{II}}_{k}\}_{k\in{\cal K}_m}}~ \sum_{k\in{\cal K}_m}\Big|
 \frac{\alpha^{\text{II}}_{k}\beta^{\text{II}}_m g_m h_{m,k}}{\eta^{\text{II}}}  -1\Big|^2\delta_{k}^2 + \sum_{m^\prime\in{\cal M}\setminus\{m\}} \sum_{k\in{\cal K}_m} \frac{|\alpha^{\text{II}}_k|^2|\beta^{\text{II}}_{m^\prime}|^2|g_{m^\prime}|^2|h_{m^\prime,k}|^2
 \delta_k^2}{(\eta^{\text{II}})^2} \\
 &~~~{\rm s. t.}~|\alpha^{\text{II}}_{k}|^2\delta_{k}^2\leq P_{k},~\forall k\in{\cal K}_m\\
 &~~~~~~ \sum_{k\in{\cal K}_m}|\alpha^{\text{II}}_{k}|^2|h_{m,k}|^2\delta_{k}^2 \leq C_m,
 \end{align}
 \end{subequations}
 where $C_m\triangleq \frac{P_{R,m}}{|\beta^{\text{II}}_m|^2} - \sum_{m^\prime\in{\cal M}\setminus\{m\}}\sum_{i\in{\cal K}_{m^\prime}} |\alpha^{\text{II}}_i|^2|h_{m,i}|^2\delta_i^2 - \sigma_{m}^2$. Denote by $\{\alpha_k^{\text{II*}}\}_{k\in{\cal K}_m}$ the optimal solution to problem~(P3.$m$), which will be obtained in Section~IV-B.

 Next, with the optimized transmit coefficients of the $K$ WDs by (P3.$m$) with $m\in{\cal M}$, the FC needs to jointly optimize its de-noising factor $\eta^{\text{II}}$ and the AF coefficients $\{\beta_m^{\text{II}}\}_{m\in{\cal M}}$ of the $M$ relays. The information required for the FC to make decision includes the channel magnitude information $\{|g_m|\}_{m\in{\cal M}}$ and $\{|h_{m,k}|\}_{m\in{\cal M},k\in\bigcup_{m\in{\cal M}}{\cal K}_m}$. Accordingly, the FC needs to solve the following problem:
 \begin{subequations}
 \begin{align}
 ({\rm P4}):& \min_{\bm \beta^{\text{II}},\eta^{\text{II}}} \sum_{m\in{\cal M}} \Bigg[\sum_{k\in{\cal K}_m}\Big|
 \frac{\beta^{\text{II}}_m \alpha^{\text{II}}_{k}|g_m h_{m,k}|}{\eta^{\text{II}}} -1\Big|^2\delta_{k}^2 + \sum_{m^\prime\in{\cal M}\setminus\{m\}} \sum_{k\in{\cal K}_m} \frac{|\alpha^{\text{II}}_k|^2|\beta^{\text{II}}_{m^\prime}|^2|g_{m^\prime}|^2|h_{m^\prime,k}|^2
 \delta_k^2}{(\eta^{\text{II}})^2} \Bigg] \notag \\
 &\quad\quad\quad +  \frac{\sum_{m\in{\cal M}}|\beta^{\text{II}}_m|^2|g_m|^2\sigma_m^2+ \sigma_0^2}{(\eta^{\text{II}})^2}  \\
 &{\rm s. t.}~~\eta^{\text{II}}>0~{\text{and}}~|\beta^{\text{II}}_m|^2 \leq \frac{P_{R,m}}{D_m},~\forall m\in{\cal M},
 \end{align}
 \end{subequations}
 where $D_m\triangleq \sum_{k\in{\cal K}_m}|\alpha^{\text{II}}_{k}|^2|h_{m,k}|^2\delta_{k}^2 + \sum_{m^\prime\in{\cal M}\setminus\{m\}} \sum_{i\in{\cal K}_{m^\prime}} |\alpha^{\text{II}}_i|^2|h_{m,i}|^2\delta_i^2 + \sigma_{m}^2$. Denote by $(\bm\beta^{\text{II*}},\eta^{\text{II}*})$ the optimal solution to problem~(P4), which will be obtained in Section~IV-C.

 The implementation for the proposed decentralized algorithm design is described as follows. In each iteration, at the lower layer, each relay $m\in{\cal M}$ takes turns to update its associated WDs' transmit coefficients $\{\alpha_k^{\text{II}}\}_{k\in{\cal K}_m}$ by solving problem (P3.$m$), and then shares the updated values of $\{\alpha_k^{\text{II}}\}_{k\in{\cal K}_m}$ to the other $(M-1)$ relays via the inter-relay communication links. Next, at the higher layer, the FC coordinates the $M$ relays to solve problem (P4). These operations are iterated until convergence.

 In the following, we focus on solving problems (P3.$m$) and (P4), respectively.

\subsection{Optimal Solution of Problem (P3.$m$)}
 Notice that problem (P3.$m$) is convex and satisfies the Slater's condition\cite{Boyd_book}. Therefore, strong duality holds for (P3.$m$). Let $\alpha_k^{\text{II*}}=\bar{\alpha}_k^{\text{II*}}e^{j\angle{\alpha_k^{\text{II*}}}}$, where $\bar{\alpha}_k^{\text{II*}}=|\alpha_k^{\text{II*}}|$ and $\angle{\alpha_k^{\text{II*}}}$ denote the magnitude and phase of $\alpha_k^{\text{II*}}$, respectively. We then have the follow lemma.

 \begin{lemma}\label{lemma.wd-tx-phase-new}
 At the optimality of (P3.$m$), it must hold that
 \begin{align}\label{lem.eq-phase-pCSI}
 \angle{\alpha_k^{\text{II*}}}  = -\angle{\beta^{\text{II}}_m g_m h_{m,k}},~\forall k\in{\cal K}_m.
 \end{align}
 \end{lemma}
 \begin{IEEEproof}
 See Appendix~\ref{Proof-lemma.wd-tx-phase-new}.
 \end{IEEEproof}

 Lemma~\ref{lemma.wd-tx-phase-new} indicates that at the optimality of problem (P3.$m$), the phase $\angle{\alpha_k^{\text{II*}}}$ of the transmit coefficient of each WD $k\in{\cal M}$ is opposite to that of the effective WD-relay-FC channel coefficient, $\beta_m^{\text{II}}g_mh_{m,k}$, of WD $k$, which is only related to the WD $k$'s associated relay $m$. This is significantly different from the case with global CSI, where the phase  $\angle\alpha^{\text{I*}}_{k} = -\angle{\bm h_k^T\bm\Lambda_g{\bm \beta^{\text{I}}}}$ (c.f. \eqref{eq.alpha_phase_I} in Lemma~\ref{prop.alpha}) of the transmit coefficient of each WD $k\in{\cal K}_m$ is determined by the channels associated with {\em all} the $M$ relays. This shows that the design in \eqref{lem.eq-phase-pCSI} significantly reduces the CSI requirements and thus signaling overhead.

 Furthermore, building on the KKT optimality conditions, we are ready to obtain the closed-formed solution of $\{\bar{\alpha}_k^{\text{II*}}\}_{k\in{\cal K}_m}$.

 \begin{proposition}\label{prop.alpha-wdk}
 The optimal solution $\{\bar{\alpha}_k^{\text{II*}}\}_{k\in{\cal K}_m}$ of problem~(P3.$m$) is expressed as
 \begin{align*}
 &\bar{\alpha}^{\text{II*}}_k =\min\Bigg\{ \frac{\frac{|\beta^{\text{II}}_mg_mh_{m,k}|}{\eta^{\text{II}}}}{\frac{|\beta^{\text{II}}_mg_mh_{m,k}|^2+\Delta_{k,m}^2}{(\eta^{\text{II}})^2}
 +\nu_m^{\rm *}|h_{m,k}|^2},\sqrt{\frac{P_k}{\delta_k^2}}\Bigg\},~\forall k\in{\cal K}_m,
 \end{align*}
 where $\Delta_{k,m}^2\triangleq \sum_{m^\prime\in{\cal M}\setminus\{m\}}|\beta^{\text{II}}_{m^\prime}g_{m^{\prime}}h_{m^{\prime},k}|^2$, $\forall k\in{\cal K}_m$, and the optimal Lagrange multiplier $\nu_m^{*}\geq 0$ satisfies the complementary slackness condition of
 \begin{align}\label{eq-mu-m-opt}
 \nu_m^{*}\Big(\sum_{k\in{\cal K}_m}(\bar{\alpha}^{\text{II*}}_k)^2|h_{m,k}|^2\delta_k^2 - C_m\Big) = 0.
 \end{align}
 Here, the optimal Lagrange multiplier $\nu_m^{*}$ can be obtained via bisection based on \eqref{eq-mu-m-opt}.
 \end{proposition}
 \begin{IEEEproof}
 The proof is similar to that of Proposition~\ref{prop.alpha-value}, and we omit it for brevity. 
 \end{IEEEproof}

 Proposition~\ref{prop.alpha-wdk} reveals a regularized channel-inversion structure for the magnitude $\bar{\alpha}^{\text{II*}}_k$ of the transmit coefficient of each WD $k\in{\cal K}_m$ at the optimality of problem (P3.$m$), where the effective channel coefficient is $\beta^{\text{II}}_mg_mh_{m,k}$. Different from the optimal magnitude $\bar{\alpha}^{\text{I*}}_k$ (c.f.~\eqref{eq.alpha_magnitude} in Proposition~\ref{prop.alpha-value}) in the case with global CSI, the regularization component for $\bar{\alpha}^{\text{II*}}_k$ in the case with partial CSI {\em additionally} includes the inter-relay-interference-induce error $\sum_{m^\prime\in{\cal M}\setminus\{m\}}|\beta^{\text{II}}_{m^\prime}g_{m^{\prime}}h_{m^{\prime},k}|^2$.

 By combining Lemma~\ref{lemma.wd-tx-phase-new} and Proposition~\ref{prop.alpha-wdk}, the optimal solution $\{\alpha^{\text{II*}}_k\}_{k\in{\cal K}_m}$ to problem~(P3.$m$) is finally obtained as
 \begin{align}\label{eq.sol-alpha-distr}
 \alpha^{\text{II*}}_k = \bar{\alpha}^{\text{II*}}_k e^{-j\angle{\beta^{\text{II}}_m g_m h_{m,k}}},~\forall k\in{\cal K}_m.
 \end{align}

 \subsection{Optimal Solution to Problem (P4)}
 Next, we consider that problem (P4) for the FC and relays, which is a non-convex problem due to the coupling of design variables $\bm \beta^{\text{II}}$ and $\eta^{\text{II}}$. To deal with this issue, we define ${\gamma^{\text{II}}}\triangleq \frac{1}{(\eta^{\text{II}})^2}$ and ${\tilde{\beta}_m^{\text{II}}} \triangleq \frac{\beta^{\text{II}}_m}{\eta^{\text{II}}}$, $\forall m\in{\cal M}$. By replacing $\frac{1}{(\eta^{\text{II}})^2}$ and $\frac{\beta^{\text{II}}_m}{\eta^{\text{II}}}$ with ${\gamma^{\text{II}}}$ and ${\tilde{\beta}_m^{\text{II}}}$, respectively, the MSE minimization problem~(P4) is equivalently recast as
 \begin{subequations}\label{eq.prob-q-eta-new}
 \begin{align}
 &\min_{\tilde{\bm \beta}^{\text{II}},{\gamma^{\text{II}}}}~
 \sum_{m\in{\cal M}} \Big( \sum_{k\in{\cal K}_m}\Big|
  {\tilde{\beta}_m^{\text{II}}} \alpha^{\text{II}}_k |g_m h_{m,k}|   - 1 \Big|^2\delta_{k}^2 + |{\tilde{\beta}_m^{\text{II}}}|^2 |g_m|^2\sigma_m^2 \notag \\
 &~~~~~~~~~~~~~~~~~~ + \sum_{m^\prime\in{\cal M}\setminus\{m\}}\sum_{i\in{\cal K}_{m^\prime}} |\alpha^{\text{II}}_i|^2|\tilde{\beta}_{m}^{\text{II}}|^2|g_{m}|^2|h_{m^\prime,i}|^2\delta_i^2  \Big)
 +{\gamma^{\text{II}}}\sigma_0^2  \\
 &{\rm s. t.}~~A_m |{\tilde{\beta}_m^{\text{II}}}|^2 \leq {\gamma^{\text{II}}} P_{R,m},~\forall m\in{\cal M}\\
 &\quad\quad {\gamma^{\text{II}}}> 0,
 \end{align}
 \end{subequations}
 where $A_m \triangleq \sum_{k\in{\cal K}_m}|\alpha^{\text{II}}_{k}|^2|h_{m,k}|^2\delta_{k}^2+\sum_{m^\prime\in{\cal M}\setminus\{m\}} \sum_{i\in{\cal K}_{m^\prime}}|\alpha^{\text{II}} _i|^2|h_{m^\prime,i}|^2\delta_i^2 + \sigma_{m}^2$, $\forall m\in{\cal M}$, and $\tilde{\bm \beta}^{\text{II}} \triangleq [\tilde{\beta}^{\text{II}}_1,...,\tilde{\beta}^{\text{II}}_M]^T$. It is clear that \eqref{eq.prob-q-eta-new} is a convex optimization problem and satisfies the Slater's condition\cite{Boyd_book}. Thus, problem \eqref{eq.prob-q-eta-new} has a zero duality gap. Denote by $(\tilde{\bm \beta}^{\text{II*}},{\gamma^{\text{II*}}})$ the optimal solution to \eqref{eq.prob-q-eta-new} and denote by $\bm \omega^{\rm *}\triangleq[\omega_1^*,...,\omega_M^*]^T\in\mathbb{R}^{M\times 1}$ the optimal Lagrange multiplier vector associated with the constraints in (\ref{eq.prob-q-eta-new}b). We thus have the following proposition.

 \begin{proposition}\label{prop.opt-q}
 At the optimality of problem~\eqref{eq.prob-q-eta-new}, the optimal primal-dual pair $(\tilde{\bm \beta}^{\text{II*}},{\gamma^{\text{II*}}},\bm \omega^{\rm *})$ satisfies that
 \begin{subequations}\label{eq.prob-sol-II}
 \begin{align}
 &{\tilde{\beta}_m^{\text{II*}}} = \begin{cases} \frac{\sum_{k\in{\cal K}_m} (\alpha^{\text{II}}_k)^H|h_{m,k}|\delta_k^2}
 {\big(\sum_{k\in{\cal K}_m}|\alpha^{\text{II}}_k|^2|h_{m,k}|^2\delta_k^2
 +\sigma_m^2+\breve{\cal I}_m\big)|g_m| }, &{\rm if}~m\neq m^{*} \\
 \frac{\sum_{k\in{\cal K}_m} (\alpha^{{\text{II}}}_k)^H |h_{m,k}|\delta_k^2}
 {\big(\sum_{k\in{\cal K}_m}|\alpha^{\text{II}}_k|^2|h_{m,k}|^2\delta_k^2
 +\sigma_m^2+\breve{\cal I}_m\big)|g_m| + \frac{A_m\sigma_0^2}{|g_m|P_{R,m}} }, &{\rm if}~m = m^{*}
  \end{cases} \\
 &{\gamma^{\text{II}}}^{\rm *} = \frac{A_{m^{\rm *}}|\tilde{\beta}^{\text{II*}}_{m^{\rm *}}|^2}{P_{R,{m^{*}}}}\\
 &\omega_m^{*} = \begin{cases}
 0,&{\rm if}~m\neq m^{*}\\
 \frac{\sigma_0^2}{P_{R,m}},&{\rm if}~m=m^{*},
 \end{cases}
 \end{align}
 \end{subequations}
 where $m^{\rm *} \triangleq
 \argmax_{m\in{\cal M}}~ \frac{A_m|\tilde{\beta}^{\text{II*}}_m|^2}{P_{R,m}}$ and $\breve{\cal I}_m\triangleq \sum_{m^\prime\in{\cal M}\setminus\{m\}} \sum_{i\in{\cal K}_{m^\prime}}|\alpha^{\text{II}} _i|^2|h_{m^\prime,i}|^2\delta_i^2 + \sigma_{m}^2$, $\forall m\in{\cal M}$.
 \end{proposition}

 \begin{IEEEproof}
 The proof is similar to that of Proposition~\ref{Prop.p3-1}, and we omit it for brevity.
 \end{IEEEproof}

 Proposition~\ref{Prop.p3-1} indicates that the optimal $\tilde{\beta}^{\text{II}*}_m$ has a scaled channel-magnitude-inversion structure for each relay $m\in{\cal M}$ with respect to the relay channel magnitude $|g_m|$, and the scaling term $\frac{\sum_{k\in{\cal K}_m} (\alpha^{\text{II}}_k)^H|h_{m,k}|\delta_k^2}
 {\sum_{k\in{\cal K}_m}|\alpha^{\text{II}}_k|^2|h_{m,k}|^2\delta_k^2
 +\sigma_m^2+\breve{\cal I}_m}$ has a MMSE-like structure. This implies that each relay $m\in{\cal M}$ needs to balance the tradeoff between maximizing the signal power of its associated WDs and minimizing the sum of the inter-relay-interference and noise power (i.e., $\breve{\cal I}_m +\sigma_m^2$). This is different from the case with global CSI, where the FC can design $\tilde{\beta}_m^{\text{I}}$ to cancel the inter-relay interference by leveraging the global channel phase related information. In addition, Proposition~\ref{Prop.p3-1} also reveals that in general at least one dual variable $\omega_{m^*}^{*}=\frac{\sigma_0^2}{P_{R,m^*}}$ needs to be non-zero, which means that relay $m^*$ utilizes its full power in AF relaying; by contrast, the other $(M-1)$ relays adopt the channel inversion power control without violating their power constraints. This is similar to the case with global CSI.

 Similarly to Proposition~\ref{Prop.p3-1}, we can first employ the exhaustive search or bisection search to find the optimal relay index $m^*$, and then obtain $(\tilde{\bm \beta}^{\text{II*}},{\gamma^{\text{II*}}})$ according to (\ref{eq.prob-sol-II}a) and (\ref{eq.prob-sol-II}b). With $(\tilde{\bm \beta}^{\text{II*}},{\gamma^{\text{II*}}})$ obtained, the optimal solution $(\bm\beta^{\text{II*}},\eta^{\text{II*}})$ to problem~(P4) is obtained by
 \begin{subequations}\label{eq.sol-beta-eta-distr}
 \begin{align}
 &\eta^{\text{II*}} = \frac{1}{\sqrt{{\gamma^{\text{II*}}}}}, \\
 &\bm \beta^{\text{II*}} = \eta^{\text{II*}}\tilde{\bm \beta}^{\text{II*}}.
 \end{align}
 \end{subequations}

\subsection{Complete Decentralized Algorithm for Solving (P2)}
 By combining Propositions \ref{prop.alpha-wdk} and \ref{prop.opt-q} together with \eqref{eq.sol-alpha-distr} and \eqref{eq.sol-beta-eta-distr}, the decentralized algorithm for solving problem~(P2) with partial CSI is summarized as Algorithm~2 in Table~II. We use ${\tt MSE}^{\text{II}(n)}$ to denote ${\tt MSE}^{\text{II}}(\bm\alpha^{\text{II}(n)},\bm \beta^{\text{II}(n)}, \eta^{\text{II}(n)})$. As problems (P3.$m$) and (P4) are optimally solved, $\forall m\in{\cal M}$, it is ensured that ${\tt MSE}^{\text{II}(n)}$ generated by Algorithm~2 is non-increasing over iterations. Therefore, the convergence of the proposed decentralized Algorithm~2 is guaranteed.


\begin{table}[htp]
\begin{center}
\caption{Proposed Decentralized Algorithm 2 for Solving Problem (P2) with Partial CSI}
\hrule
\begin{itemize}
\item[a)] {\bf Initialization:} $n=1$, $\epsilon>0$, $|\alpha_{k}^{\text{II}(0)}|^2\leq P_{k}$, $\forall k\in{\cal K}_m$, $|\beta^{\text{II}(0)}_m|^2\big(\sum_{m^\prime\in{\cal M}}\sum_{k\in{\cal K}_m}|\alpha^{\text{II}(0)}_k h_{k,m}|^2+\sigma_{R,m}^2\big) \leq P_{R,m}$, the inter-relay interference values $\sum_{i\in{\cal K}_{m^\prime}}|\alpha^{\text{II}(0)}_{i}|^2|h_{m,i}|^2\delta^2_{i}$, $\forall m^\prime\in{\cal M}\setminus\{m\}$, $m\in{\cal M}$.
\item[b)] {\bf While} $({\tt MSE}^{\text{II}(n-1)}-{\tt MSE}^{\text{II}(n)})/{\tt MSE}^{\text{II}(n-1)}>\epsilon$ {\bf do}
 \begin{itemize}
 \item {\bf For} $m=1,...,M$, {\bf do}
 \begin{itemize}
 \item Under given $(\beta_1^{\text{II}(n-1)},...,\beta_M^{\text{II}(n-1)},
     \eta^{\text{II}(n-1)})$, each relay $m$ obtains the optimal solution $\{\alpha_{k}^{\text{II}(n)}\}_{k\in{\cal K}_m}$ to problem~(P3.$m$) based on Proposition~\ref{prop.alpha-wdk} and~\eqref{eq.sol-alpha-distr};
 \item Each relay $m\in{\cal M}$ shares the updated transmit coefficients $\{\alpha_{k}^{\text{II}(n)}\}_{k\in{\cal K}_m}$ to the other $(M-1)$ relays via inter-relay communication links;
 \end{itemize}
 \item {\bf End for}
\item Under given $(\alpha_{1}^{\text{II}(n)},...,\alpha_{K}^{\text{II}(n)})$, the FC obtains the optimal solution $(\{\beta_m^{\text{II}(n)}\}_{m\in{\cal M}},\eta^{\text{II}(n)})$ for problem~(P4) based on Proposition~\ref{prop.opt-q} and~\eqref{eq.sol-beta-eta-distr};
 \item $n=n+1$;
    \end{itemize}
\item[c)] {\bf End while}
\item[e)] {\bf Output:} $(\{\alpha_k^{\text{II}(n)}\}_{k\in{\cal K}_m,m\in{\cal M}},\{\beta_m^{\text{II}(n)}
    \}_{m\in{\cal M}},
    \eta^{\text{II}(n)})$ for problem (P2) with partial CSI.
\end{itemize}
\hrule
\end{center}
\end{table}

\section{Numerical Results}\label{sec:numerical}
 In this section, we provide numerical results to evaluate the performance of the proposed centralized and decentralized designs for hierarchical AirComp systems. In the simulation, we consider the distance-dependent Rayleigh fading channel models\cite{Goldsmith05}, in which we set $h_{k,m}=\sqrt{\Omega_0d_{k,m}^{-\kappa}}h_0$ and $g_{m}=\sqrt{\Omega_0d_{m}^{-\kappa}}g_0$, where $d_{k,m}\in{\cal U}[30, 150]$ meters and $d_{m}=200$ meters denote the distances from WD $k$ to relay $m$ and from relay $m$ to the FC, respectively, $\Omega_0=-37$~dB corresponds to the path loss at a reference distance of one meter, $\kappa=3.5$ is the pathloss exponent, and $h_0\sim{\cal CN}(0,1)$ and $g_0\sim{\cal CN}(0,1)$ account for small-scale channel fading. The WD $k$ is associated with the nearest relay $m\in{\cal M}$, i.e., $k\in{\cal K}_m$ with $d_{k,m} = \min_{i\in{\cal M}}d_{k,i}$. Unless stated otherwise, we set the variances of WDs' transmit messages to be $\delta^2_k=2$, the maximum transmit power values of WDs and relays to be $P_{k}=200$ milliwatt (mW) and $P_{R,m}=800$~mW, respectively, and the noise power to be $\sigma_{m}^2=\sigma_0^2=10^{-7}$ mW, $\forall m\in{\cal M}$. The numerical results are obtained by averaging over $10^3$ randomized channel realizations. For performance comparison, we consider the following three benchmark schemes.
\begin{itemize}
\item {\em Full power transmission at WDs and relays}: In this scheme, each WD $k\in{\cal K}$ and each relay $m\in{\cal M}$ use their full transmit power, respectively. This scheme corresponds to solving problem (P1) by setting $|\alpha_k|^2\delta_k^2=P_k$, $\forall k\in{\cal K}_m$, and $|\beta^{\text{I}}_m|^2(\sum_{m^\prime\in{\cal M}}\sum_{k\in{\cal K}_{m^\prime}}|\alpha^{\text{I}}_k|^2 |h_{k,m}|^2\delta_k^2+\sigma_{m}^2) = P_{R,m}$, $\forall m\in{\cal M}$.
\item {\em Full power transmission at WDs only}: In this scheme, each WD $k\in{\cal K}$ uses the full transmit power to send data, which corresponds to solving problem (P1) by setting $|\alpha^{\text{I}}_k|^2\delta_k^2=P_k$, $\forall k\in{\cal K}_m,m\in{\cal M}$.
\item {\em Full power transmission at relays only}: In this scheme, each relay $m\in{\cal M}$ employs the full transmit power to amplify and forward its received signal to the FC, which corresponds to solving problem (P1) by setting $|\beta^{\text{I}}_m|^2(\sum_{m^\prime\in{\cal M}}\sum_{k\in{\cal K}_{m^\prime}}|\alpha^{\text{I}}_k|^2 |h_{k,m}|^2\delta_k^2+\sigma_{m}^2) = P_{R,m}$, $\forall m\in{\cal M}$.
\end{itemize}

\begin{figure}
  \centering
  \includegraphics[width = 3.5in]{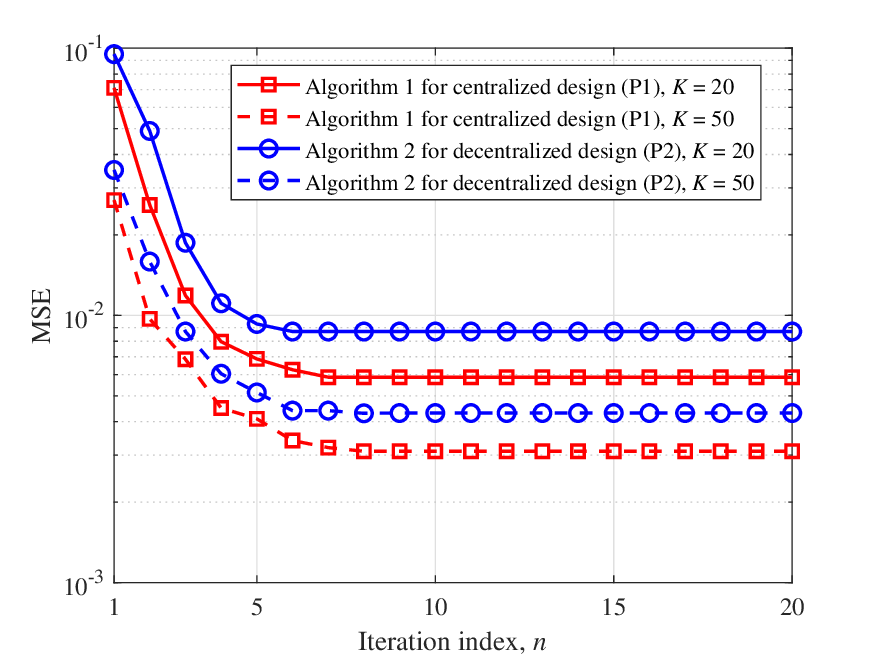}
 \caption{The convergence of the proposed centralized and decentralized algorithms, where the relay number is set as $M=5$.} \label{fig.Convergence}
 \end{figure}

 Fig.~\ref{fig.Convergence} illustrates the convergence performance of the proposed Algorithm~1 for the centralized design problem~(P1) with global CSI and Algorithm~2 for the decentralized design problem~(P2) with partial CSI under one channel realization, where the number of relays is $M=5$ and the error tolerance is set as $\epsilon=10^{-4}$. It is observed that with the number of WDs being $K=20$ and $K=50$, the proposed Algorithms~1 and 2 both converge within around 10 iterations.

 \begin{figure}
 \centering
 \begin{minipage}{0.49\textwidth}
 \centering
 \includegraphics[width = 3.2in]{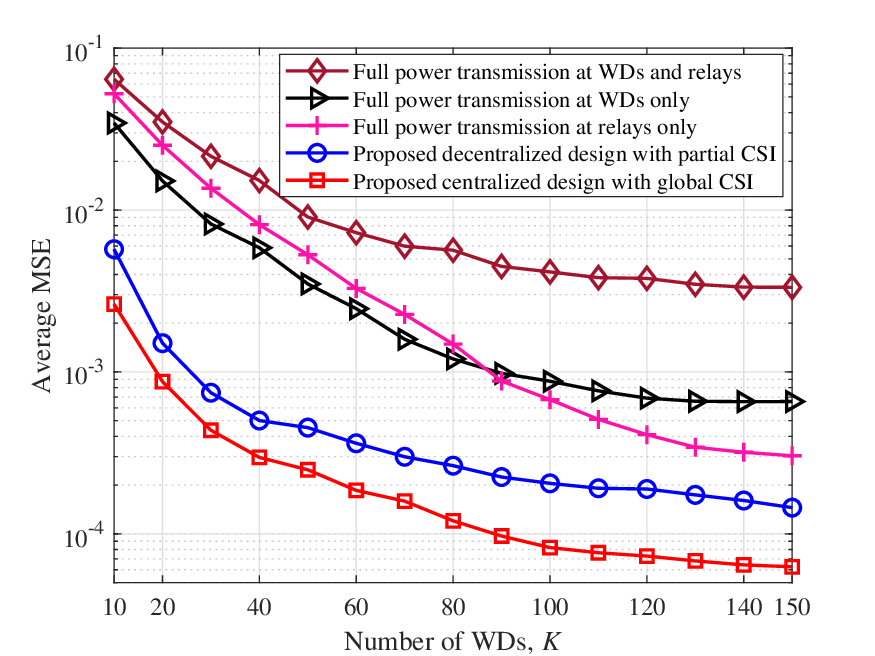}
 \caption{The average MSE performance versus the number of WDs $K$.} \label{fig.vs_K}
 \end{minipage}
 \begin{minipage}{0.49\textwidth}
 \centering
 \includegraphics[width = 3.2in]{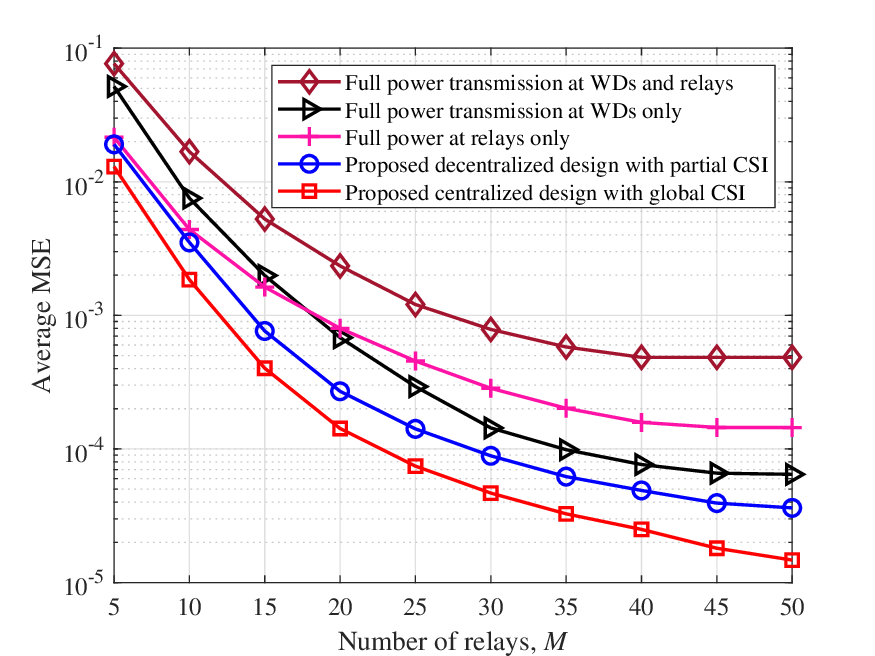}
 \caption{The average MSE performance versus the number of relays $M$.} \label{fig.vs_M}
 \end{minipage}
 \end{figure}

 Fig.~\ref{fig.vs_K} shows the average computational MSE versus the number of WDs $K$, where $M=10$ and each WD is set to be associated with the nearest relay. It is observed that as $K$ increases, the MSE value decreases for all the five schemes, and the proposed two designs outperform the three benchmark schemes in all regimes of $K$. As $K$ increases, the proposed centralized design is observed to achieve a more substantial performance gain over the proposed decentralized design, which is because the inter-relay interference becomes more severe in this case. When $K$ is small (e.g., $K\leq 40$), the full-power-transmission-at-WDs-only scheme is observed to outperform the full-power-transmission-at-relays-only scheme, but it is not true as $K$ grows large. This implies the importance of power control at WDs to reduce the signal-magnitude-misalignment-induced error as $K$ increases. The full-power-transmission-at-relays-only scheme is observed to achieve a performance close to the proposed designs at large $K$ values, which shows the benefit of full-power operation at relays in reducing the MSE in this case. Due to loss of degrees of freedom in transmit power control, the full-power-transmission-at-WDs-and-relays scheme is observed to perform inferior to the other schemes.


 Fig.~\ref{fig.vs_M} depicts the average computational MSE versus the number of relays $M$. It is observed that the MSE performance of the five schemes decreases as $M$ increases, and the proposed two designs outperform the three benchmark schemes. As the number of relays $M$ increases, the proposed centralized design is observed to achieve an increasing MSE gain over the proposed decentralized design. When $M$ is small (e.g., $M\leq 17$), the full-power-transmission-at-relays-only scheme is observed to outperform the full-power-transmission-at-WDs-only one, but the reverse holds true as $M$ grows larger in this setup. This is because the full-power-transmission-at-WDs-only scheme can adapt the transmit coefficients at relays to reduce the signal-magnitude-misalignment-induced error. The full-power-transmission-at-WDs-only scheme is observed to achieve a performance close to the proposed designs as $M$ increases. This implies the benefit for the WDs to employ full power transmission to reduce the computational MSE in the cases with a large number of relays.

 \begin{figure}
  \centering
  \begin{minipage}{0.49\textwidth}
  \includegraphics[width = 3.2in]{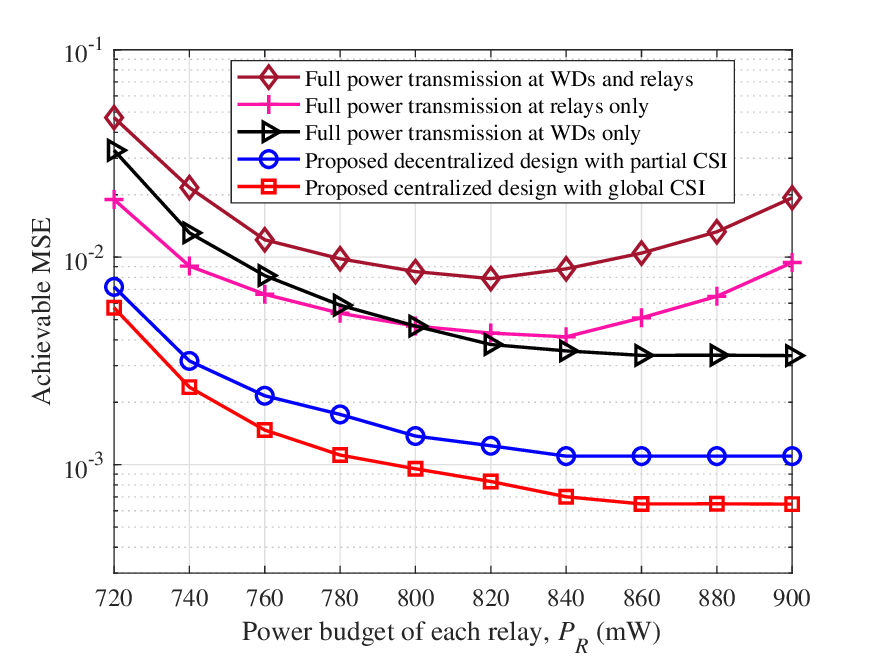}
 \caption{The average MSE performance versus the relay power budget $P_R$.} \label{fig.vs_PR}
 \end{minipage}
 \begin{minipage}{0.49\textwidth}
  \centering
  \includegraphics[width = 3.2in]{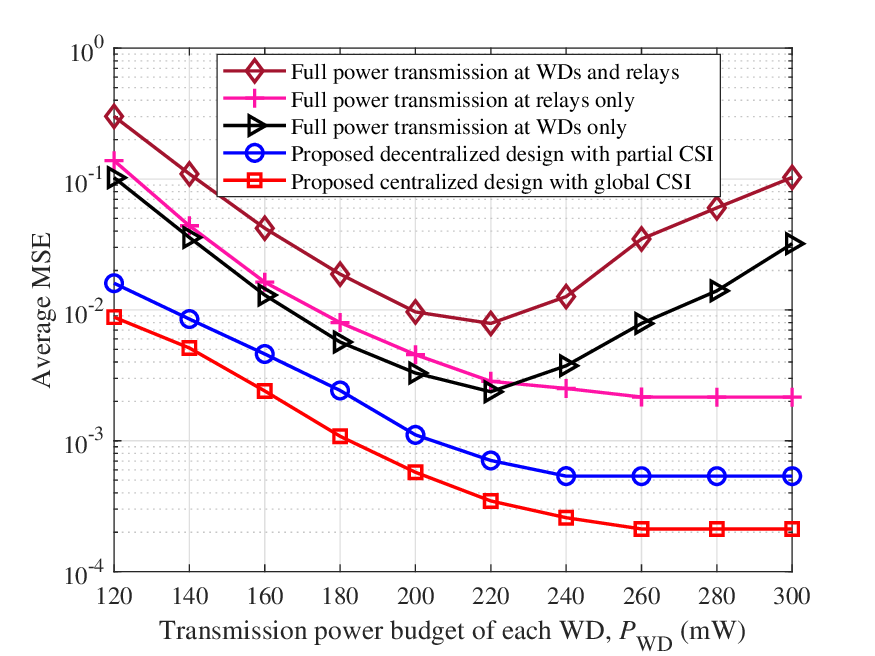}
 \caption{The average MSE performance versus the WD power budget $P_{\rm WD}$.} \label{fig.vs_Pk}
 \end{minipage}
 \end{figure}

 Fig.~\ref{fig.vs_PR} shows the average computational MSE versus the power budget of each relay $P_R$, where $K=30$, $M=5$, and we set $P_{R,m}=P_R$, $\forall m\in{\cal M}$. The proposed two designs are observed to outperform the three benchmark schemes. As $P_R$ increases, the MSE value is observed to decrease for the proposed designs and the full-power-transmission-at-WDs-only scheme. This is because these three schemes can freely adjust the relays' AF coefficients, which is benefited from the increasing power budget for the AF procedure from the relays to the AP. By contrast, as $P_R$ increases, it is observed that the full-power-transmission-at-relays-only scheme and full-power-transmission-at-WDs-and-relays scheme first harvest the benefit of a larger power budget of the relays, but then suffer from a significant MSE performance loss. This is because a large signal-magnitude-misalignment-induced error is incurred by always letting each relay utilize the sufficiently large power budget in AF relaying.


 Fig.~\ref{fig.vs_Pk} shows the average computational MSE versus the transmission power budget of each WD $P_{\rm WD}$, where $K=30$, $M=5$, and we set $P_{k}=P_{\rm WD}$, $\forall k\in{\cal K}$. Again, the proposed decentralized and centralized schemes are observed to outperform the three benchmark schemes. As $P_{\rm WD}$ increases, the MSE performance achieved by the proposed two designs and the full-power-transmission-at-relays-only scheme is expected to decrease. This is due to the limited degree of design freedom in adjusting the WDs' transmit coefficients. For the full-power-transmission-at-WDs-only scheme and full-power-transmission-at-WDs-and-relays scheme, it is observed that the achieved MSE value first decreases and then increases as $P_{\rm WD}$ decreases. This is because a large signal-magnitude-misalignment-induced error is incurred by always setting each WD's transmission at its full power.

\section{Conclusion}\label{sec:Conclusion}
 In this paper, we presented the hierarchical Aircomp designs with AF relaying to minimize the computational MSE at the FC subject the individual power constraints of the WDs and relays, by jointly optimizing the transmit coefficients of the WDs, the AF coefficients of the relays, and the de-noising factor of the FC. We developed the alternating-optimization-based solution for the centralized design with global CSI, and showed the transmit power at each WD/relay follows a new regularized composite-channel-inversion structure to strike a balance between minimizing the signal-magnitude-misalignment-induced error and the noise-induced error. To reduce the signaling overhead in obtaining global CSI, we proposed a decentralized design with partial CSI, which facilitates the decentralized implementation for hierarchical AirComp systems without inter-relay coordination at lower layer designs. Numerical results showed the fast convergence and the MSE performance gains of the proposed designs as compared to the benchmark schemes without the joint optimization. This work is expected to motivate several interesting research directions on hierarchical AirComp designs with multi-antenna and/or multi-hop setups, WD-relay associations, and/or with multiple functions to be computed.

%

\appendix

\subsection{Proof of Lemma~\ref{prop.alpha}}\label{prop.alpha-proof}

 We prove Lemma~\ref{prop.alpha} by contradiction. Suppose that the optimal solution to problem~\eqref{eq.prob-p2} is given by $\bm\alpha^{\rm I\prime}$, where $\alpha^{\rm I\prime}_k = |\alpha^{\text{I*}}_k|e^{j\phi_k^{\prime}}$, $\forall k\in{\cal K}$. For $\bm \alpha^{\rm I\prime}$, there exists one WD $k_0\in{\cal K}$ and $0<\phi_0<2\pi$, such that $\phi_{k_0}^{\prime}= -\angle{\bm h_{k_0}^T\bm\Lambda_g{\bm \beta^{\text{I}}}}+\phi_0$ and $\phi_{k}^{\prime} = -\angle{\bm h_k^T\bm\Lambda_g{\bm \beta^{\text{I}}}}$, $\forall k\in{\cal K}\setminus\{k_0\}$. We can always construct a feasible solution $\bm\alpha^{\rm I\prime\prime}$, such that $\alpha^{\rm I\prime\prime}_{k_0}=|\alpha^{\text{I*}}_{k_0}|e^{j\phi_{k_0}^{\prime\prime}}$ with $\phi_{k_0}^{\prime\prime}= -\angle{\bm h_k^T\bm\Lambda_g{\bm \beta^{\text{I}}}}$, and $\alpha^{\rm I\prime\prime}_{k}=\alpha^{\rm I\prime}_{k}$, $\forall k\in{\cal K}\setminus\{k_0\}$. The gap $\Delta$ between the MSEs achieved by $\bm \alpha^{\rm I\prime}$ and $\bm \alpha^{\rm I\prime\prime}$ is then
 \begin{align}
 \Delta &= \sum_{m\in{\cal M}}\sum_{k\in{\cal K}_m}\Big|\frac{\alpha^{\rm I\prime}_{k}\bm h_k^T\bm\Lambda_g{\bm \beta^{\text{I}}}}{\eta^{\text{I}}}-1\Big|^2\delta_k^2 - \sum_{m\in{\cal M}}\sum_{k\in{\cal K}_m}\Big|\frac{\alpha^{\rm I\prime\prime}_{k}\bm h_k^T\bm\Lambda_g{\bm \beta^{\text{I}}}}{\eta^{\text{I}}}-1\Big|^2\delta_k^2 \notag \\
 &= \Big|\frac{\alpha^{\rm I\prime}_{k_0}\bm h_{k_0}^T\bm\Lambda_g{\bm \beta^{\text{I}}}}{\eta^{\text{I}}}-1\Big|^2\delta_{k_0}^2 - \Big|\frac{\alpha^{\rm I\prime\prime}_{k_0}\bm h_{k_0}^T\bm\Lambda_g{\bm \beta^{\text{I}}}}{\eta^{\text{I}}}-1\Big|^2\delta_{k_0}^2 \notag \\
 &=\frac{2|\alpha^{\text{I*}}_{k_0}\bm h_{k_0}^T\bm\Lambda_g{\bm \beta^{\text{I}}}|\delta_{k_0}^2}{\eta^{\text{I}}}(1-\frac{e^{j\phi_0}+e^{-j\phi_0}}{2}) \notag \\
 &=\frac{2|\alpha^{\text{I*}}_{k_0}\bm h_{k_0}^T\bm\Lambda_g{\bm \beta^{\text{I}}}|\delta_{k_0}^2}{\eta^{\text{I}}}(1-\cos(\phi_0)).
 \end{align}
 Due to the fact that $(1-\cos(\phi_0)>0$ for $0<\phi_0<2\pi$, it holds that $\Delta>0$. The positive gap $\Delta>0$ shows that the feasible solution $\bm\alpha^{\rm I\prime\prime}$ achieves a smaller objective value than that by $\bm\alpha^{\rm I\prime}$. Hence, the assumption of $\bm\alpha^{\rm I\prime}$ being the optimal solution to problem~\eqref{eq.prob-p2} cannot be true. It thus follows that $\angle{\alpha_k^{\text{I*}}}= -\angle{\bm h_k^T\bm\Lambda_g{\bm \beta^{\text{I}}}}$, $\forall k\in{\cal K}_m,m\in{\cal M}$, which completes the proof of Lemma~\ref{prop.alpha}.


\subsection{Proof of Proposition~\ref{prop.alpha-value}}\label{prop.alpha-value-proof}
 Let $\mu_{m}\geq 0$ denote the dual variables associated with the $m$th constraint in (\ref{eq.prob-p21}c), $\forall m\in{\cal M}$. The partial Lagrangian of problem (\ref{eq.prob-p21}) is given as
 \begin{align*}
 &{\cal L}(\bm \alpha^{\text{I}},\bm\mu) \triangleq \sum_{m\in{\cal M}}\sum_{k\in{\cal K}_m} \Big(\frac{\bar{\alpha}^{\text{I}}_{k}|\bm h_k^T \bm\Lambda_g\bm\beta^{\text{I}}|}{\eta^{\text{I}}}-1\Big)^2 + \sum_{m\in{\cal M}}\mu_{m}\Big(\sum_{m^\prime\in{\cal M}}\sum_{k\in{\cal K}_{m^\prime}}(\bar{\alpha}^{\text{I}}_{k})^2|h_{m,k}|^2+\sigma^2_m-\frac{P_{R,m}}{|\beta^{\text{I}}_m|^2}\Big)\notag\\
 &\quad =  \sum_{m\in{\cal M}}\sum_{k\in{\cal K}_m}\Big[\Big(\frac{\bar{\alpha}^{\text{I}}_{k}{|{\bm h}^T_{k}\bm\Lambda_g\bm\beta^{\text{I}}|}}{{\eta}^{\text{I}}}-1\Big)^2\delta^2_k +\sum_{m^\prime\in{\cal M}}\mu_{m^\prime}(\bar{\alpha}^{\text{I}}_{k})^2|h_{m^\prime,k}|^2\delta^2_k\Big]
 +\sum_{m\in{\cal M}}\mu_{m}\Big(\sigma_m^2-\frac{P_{R,m}}{|\beta_m^{\text{I}}|^2}\Big),
\end{align*}
 where $\bm \mu\triangleq [\mu_1,...,\mu_M]^T$. The necessary and sufficient conditions for the optimal primal and dual variables are given by the KKT optimality conditions\cite{Boyd_book}, which are expressed below.
 \begin{subequations}\label{eq.KKT}
 \begin{align}
 &\bar{\alpha}^{\text{I*}}_k \leq \sqrt{P_{k}/\delta^2_k},~~\forall k\in{\cal K}_m,m\in{\cal M} \\
 &\sum_{m^\prime\in{\cal M}}\sum_{k\in{\cal K}_m}(\bar{\alpha}^{\text{I*}}_{k})^2|h_{m,j}|^2\delta^2_k + \sigma_{m}^2
 \leq \frac{P_{R,m}}{|\beta^{\text{I}}_m|^2},~\forall m\in{\cal M}\\
 &\mu^{\rm *}_m\geq 0,~\forall m\in{\cal M}\\
 &\mu^{\rm *}_{m}\Big(\sum_{m^{\prime}\in{\cal M}}\sum_{k\in{\cal K}_{m^\prime}}(\bar{\alpha}^{\text{I*}}_{k})^2|h_{m^\prime,k}|^2\delta^2_k +
 \sigma^2_m-\frac{P_{R,m}}{|\beta^{\text{I}}_m|^2}\Big)=0,~\forall m\in{\cal M} \\
 &\frac{\partial {\cal L}}{\partial \bar{\alpha}^{\text{I*}}_{k}}= 2\bar{\alpha}^{\text{I*}}_{k} \Big(\frac{|\bm h^T_{k}\bm\Lambda_g\bm\beta^{\text{I}}|^2}{(\eta^{\text{I}})^2} + \sum_{m^\prime\in{\cal M}} \mu^{*}_{m^\prime}|h_{m^\prime,k}|^2\Big)\delta^2_k - \frac{2|\bm h_{k}^T\bm\Lambda_g\bm\beta^{\text{I}}|\delta^2_k}{\eta^{\text{I}}}=0,~\forall k\in{\cal K}_m,m\in{\cal M},
 \end{align}
 \end{subequations}
 where (\ref{eq.KKT}a)-(\ref{eq.KKT}b) and (\ref{eq.KKT}c) denote the primal feasibility and dual feasibility conditions, respectively, (\ref{eq.KKT}d) denotes the complementary slackness conditions, and (\ref{eq.KKT}e) denotes the Lagrangian stationary conditions at the primal optimal point. Based on the KKT optimality conditions in \eqref{eq.KKT}, the optimal solution $\bar{\bm \alpha}^{\text{I*}}$ for problem \eqref{eq.prob-p21} is obtained as
 \begin{align}
 \bar{\alpha}^{\text{I*}}_{k} = \min\left\{ \frac{\frac{|\bm h^T_{k}\bm\Lambda_g\bm\beta^{\text{I}}|}{\eta^{\text{I}}}}{\frac{|\bm h_{k}^T\bm\Lambda_g\bm\beta^{\text{I}}|^2}{(\eta^{\text{I}})^2} + \sum_{m\in{\cal M}} \mu^{\rm *}_m|h_{m,k}|^2},\sqrt{\frac{P_{k}}{\delta^2_k}}\right\},
\end{align}
 where $k\in{\cal K}_m,m\in{\cal M}$, and the nonnegative Lagrange multipliers $\{\mu^{*}_m\}_{m\in{\cal M}}$ satisfy the complementary slackness conditions in (\ref{eq.KKT}c). Therefore, we complete the proof of Proposition~\ref{prop.alpha-value}.

\subsection{Proof of Proposition~\ref{Prop.p3-1}} \label{Proof-Prop.p3-1}
 With the nonnegative real-valued Lagrange multiplier vector $\bm \lambda=[\lambda_1,...,\lambda_M]^T$ associated with the $M$ constraints in (\ref{eq.prob-p3-1}b), the partial Lagrangian of problem \eqref{eq.prob-p3-1} is expressed as
 \begin{align}\label{eq.Lag-p3-1}
 &{\cal G}(\tilde{\bm \beta}^{\text{I}},\gamma^{\text{I}},\bm \lambda) \notag\\
 &\triangleq  \sum_{m\in{\cal M}}\sum_{k\in{\cal K}_m} \Big|\alpha^{\text{I}}_{k}\bm h_k^T\bm \Lambda_g\tilde{\bm \beta}^{\text{I}}-1\Big|^2\delta_k^2 + (\tilde{\bm \beta}^{{\text{I}}})^H\bm\Lambda^H_g\bm\Lambda_{\sigma^2}\bm\Lambda_g
 \tilde{\bm \beta}^{\text{I}} +\gamma^{\text{I}}\sigma_0^2 + \sum_{m\in{\cal M}} \lambda_m(|\tilde{\beta}^{\text{I}}_m|^2-\gamma^{\text{I}} \bar{P}_{R,m}) \notag \\
 &= (\tilde{\bm\beta}^{{\text{I}}})^H \bm \Lambda_g^H \Big( \bar{\bm H} + \bm \Lambda_{\sigma^2} + \bm \Lambda_{\lambda} \Big) \bm \Lambda_g \tilde{\bm \beta}^{\text{I}} -2 {\rm Re}\Big(\sum_{m\in{\cal M}}\sum_{k\in{\cal K}_m} \alpha^{\text{I}}_k\bm h_k^T \bm \Lambda_g\tilde{\bm \beta}^{\text{I}}\Big) +  \sum_{m\in{\cal M}}\sum_{k\in{\cal K}_m} \delta_k^2 \notag \\
 &\quad + \gamma^{\text{I}}\big(\sigma_0^2 - \sum_{m\in{\cal M}} \lambda_m\bar{P}_{R,m}\big),
 \end{align}
 where $\bar{\bm H} \triangleq \sum_{m\in{\cal M}}\sum_{k\in{\cal K}_m}\delta_k^2|\alpha^{\text{I}}_k|^2\bm h_k\bm h_k^H$ and $\bm\Lambda_{{\lambda}} \triangleq {\rm diag}\big(\frac{\lambda_1}{|g_1|^2},...,\frac{\lambda_M}{|g_M|^2}\big)$. Building on \eqref{eq.Lag-p3-1}, the KKT optimality conditions on the optimal primal-dual pair $(\tilde{\bm \beta}^{\text{I*}},\gamma^{\text{I*}},\bm\lambda^{*})$ are given by
 \begin{subequations}\label{eq.p3-1-KKT}
 \begin{align}
 & \gamma^{\text{I*}}\geq 0,~|\tilde{\beta}^{\text{I*}}_m|^2 \leq \gamma^{\text{I*}}\bar{P}_{R,m},~\forall m\in{\cal M} \\
 & \lambda_m^{*} \geq 0,~\forall m\in{\cal M} \\
 & \lambda_m^{*}(|\tilde{\beta}^{\text{I*}}_m|^2 - \gamma^{\text{I*}}\bar{P}_{R,m}) =0,~\forall m\in{\cal M} \\
 &\nabla_{\tilde{\bm\beta}^{\text{I}}}{\cal G}(\tilde{\bm \beta}^{\text{I}},\gamma^{\text{I*}},\bm\lambda^{*})\Big|_{\tilde{\bm \beta}^{\text{I}}=\tilde{\bm \beta}^{\text{I*}}} = \bm 0\\
 & \nabla_{\gamma^{\text{I}}}{\cal G}(\tilde{\bm \beta}^{\text{I*}},\gamma,\bm\lambda^{*})\Big|_{\gamma^{\text{I}} =\gamma^{\text{I*}}} = 0,
 \end{align}
 \end{subequations}
 where (\ref{eq.p3-1-KKT}a) and (\ref{eq.p3-1-KKT}b) represent the primal feasibility and dual feasibility conditions, respectively, the equalities in (\ref{eq.p3-1-KKT}c) denote the complementary slackness conditions, and  (\ref{eq.p3-1-KKT}d) and (\ref{eq.p3-1-KKT}e) denote the Lagrangian stationarity conditions with respect to $\tilde{\bm\beta}^{\text{I*}}$ and $\gamma^{\text{I*}}$, respectively.

 From (\ref{eq.p3-1-KKT}d), it follows that
 $\tilde{\bm \beta}^{\text{I*}} = \bm\Lambda(\bar{\bm H} + \bm \Lambda_{\sigma^2} + \bm\Lambda_{{\lambda}^{*}} )^{-1}\big(\sum_{m\in{\cal M}}\sum_{k\in{\cal K}_m} \alpha^{\text{I}}_k\bm h_k^T\big)^H$, where $\bm\Lambda = {\rm diag} \big(\frac{g^H_1}{|g_1|^2},...,\frac{g^H_M}{|g_M|^2}\big)$ and $\bm\Lambda_{{\lambda^{*}}} = {\rm diag}\big(\frac{\lambda^{*}_1}{|g_1|^2},...,\frac{\lambda^{*}_M}{|g_M|^2}\big)$. From (\ref{eq.p3-1-KKT}e), it follows that
 \begin{align}\label{eq.lambda-kkt}
 \sigma_0^2 - \sum_{m\in{\cal M}} \lambda^{*}_m\bar{P}_{R,m} =0.
 \end{align}

 Based on the primal feasibility conditions in (\ref{eq.p3-1-KKT}a), the optimal primal variable $\gamma^{\text{I*}}$ satisfies $\gamma^{\text{I*}} \geq \max_{m\in{\cal M}} \frac{|\tilde{\beta}^{\text{I*}}_m|^2}{\bar{P}_{R,m}}$. From the Lagrangian stationary condition \eqref{eq.lambda-kkt} with respect to $\gamma^{\text{I*}}$, there must exist at least one positive dual variable $\lambda^{*}_m$; otherwise, it cannot be true that $\sigma^2_0=0$. Therefore, by denoting $m^{*} =\argmax_{m\in{\cal M}} \frac{|\tilde{\beta}^{\text{I*}}_m|^2}{\bar{P}_{R,m}}$, the primal feasibility conditions become
 \begin{align}\label{eq.primal-fea}
 |\tilde{\beta}^{\text{I*}}_m|^2 - \gamma^{\text{I*}}\bar{P}_{R,m}
 \begin{cases}
 = 0,  &{\rm if}~m=m^{*} \\
 < 0,  &{\rm if}~m\neq m^{*},
 \end{cases}
 \end{align}
 which indicates that only one primal feasibility condition becomes activated at the optimality of problem \eqref{eq.prob-p3-1}. By jointly considering \eqref{eq.primal-fea} and the complementary slackness conditions (\ref{eq.p3-1-KKT}c), it must hold that
 \begin{align}\label{eq.opt-lambda}
 \lambda^{*}_{m} = 0,~\forall m\in{\cal M}\setminus\{m^{\rm *}\}.
 \end{align}

 Based on \eqref{eq.opt-lambda}, the optimal primal variable $\gamma^{\text{I*}}$ is expressed as $ \gamma^{\text{I*}} = \frac{|\tilde{\beta}^{\text{I*}}_{m^{*}}|^2}{\bar{P}_{R,m^{*}}}$. By substituting \eqref{eq.opt-lambda} into \eqref{eq.lambda-kkt}, we then have $\lambda_{m^{*}}^{*} = \frac{\sigma^2_0}{\bar{P}_{R,{m^*}}}$. We now complete the proof of Proposition~\ref{Prop.p3-1}.

\subsection{Proof of Lemma~\ref{lemma.wd-tx-phase-new} }\label{Proof-lemma.wd-tx-phase-new}
 We prove Lemma~\ref{lemma.wd-tx-phase-new} by contradiction. Suppose that the optimal solution to problem~(P3.$m$) is given by $\{\alpha^{\rm II\prime}_k\}_{k\in{\cal K}_m}$, where $\alpha^{\rm II\prime}_k = |\alpha^{\text{II*}}_k|e^{j\theta_k^{\prime}}$, $\forall k\in{\cal K}_m$. For $\{\alpha^{\rm II\prime}_k\}_{k\in{\cal K}_m}$, there exists one WD $k_0\in{\cal K}_m$ and $0<\theta_0<2\pi$, such that $\theta_{k_0}^{\prime}= -\angle{\beta^{\text{II}}_m g_m h_{m,k_0}}+\theta_0$ and $\theta_{k}^{\prime} = -\angle{\beta^{\text{II}}_m g_m h_{m,k}}$, $\forall k\in{\cal K}_m\setminus\{k_0\}$. We can always construct a feasible solution $\{\alpha^{\rm II\prime\prime}_k\}_{k\in{\cal K}_m}$, such that $\alpha^{\rm II\prime\prime}_{k_0}=|\alpha^{\text{II*}}_k|e^{j\theta_{k_0}^{\prime\prime}}$ with $\theta_{k_0}^{\prime\prime}= -\angle{\beta^{\text{II}}_m g_m h_{m,k_0}}$, and $\alpha^{\rm II\prime\prime}_{k}=\alpha^{\rm II\prime}_{k}$, $\forall k\in{\cal K}_m\setminus\{k_0\}$. The gap $\Delta$ between the MSEs achieved by $\{\alpha^{\rm II\prime}_k\}_{k\in{\cal K}_m}$ and $\{\alpha^{\rm II\prime\prime}_k\}_{k\in{\cal K}_m}$ is given as
 \begin{align}
\Delta &= \Psi_m(\{\alpha^{\rm II\prime}_k\}_{k\in{\cal K}_m},\bm\beta^{\text{II}},\eta^{\text{II}}) - \Psi_m(\{\alpha^{\rm II\prime\prime}_k\}_{k\in{\cal K}_m},\bm\beta^{\text{II}},\eta^{\text{II}}) \notag \\
&= \Big|\frac{\alpha^{\rm II\prime}_{k_0}\beta^{\text{II}}_m g_m h_{m,k}}{\eta^{\text{II}}}  -1\Big|^2\delta_{k_0}^2 - \Big|\frac{\alpha^{\rm II\prime\prime}_{k_0}\beta^{\text{II}}_m g_m h_{m,k}}{\eta^{\text{II}}}  - 1 \Big|^2\delta_{k_0}^2 \notag \\
&=\frac{2|\alpha^{\text{II*}}_{k_0}\beta^{\text{II}}_m g_m h_{m,k}|\delta_{k_0}^2}{\eta^{\text{II}}}(1-\cos(\theta_0)).
 \end{align}
 Due to the fact that $(1-\cos(\theta_0))>0$ for $0<\theta_0<2\pi$, it always holds that $\Delta>0$. The positive gap $\Delta>0$ shows that the feasible solution $\{\alpha^{\rm II\prime\prime}_k\}_{k\in{\cal K}_m}$ achieves a smaller objective value than that by the solution $\{\alpha^{\rm II\prime}_k\}_{k\in{\cal K}_m}$. Hence, the assumption of $\{\alpha^{\rm II\prime}_k\}_{k\in{\cal K}_m}$ being the optimal solution to problem~(P3.$m$) is not true. It thus follows that $\angle{\alpha_k^{\text{II}*}}=\theta_{k}^{*}= -\angle{\beta^{\text{II}}_m g_m h_{m,k}}$, $\forall k\in{\cal K}_m$, and we now complete the proof of Lemma~\ref{lemma.wd-tx-phase-new}.

\end{document}